# What Prediction Markets Can See

*Market Formation, Settlement Legibility, and the Geography of Tradable Uncertainty in Africa and Latin America*

Ade Adegbenro[1]

June 2026

**Abstract**

Prediction markets are usually evaluated after their contracts exist, by asking how well prices forecast outcomes. We study the prior institutional margin of market formation, asking which uncertainties become tradable contracts at all. Using an audited dataset of 6,047 Africa-topic and Latin America-topic contracts listed on Polymarket and Kalshi, we construct a coded measure of settlement legibility, the degree to which an uncertainty can be worded, sourced, and credibly resolved by third parties, and validate it on 451 units under a frozen codebook, where independent double scoring reaches ordinal reliabilities of 0.92 and 0.96 on the primary dimensions and blind human benchmarks reach 0.97 and 0.92. Using this measure, we find that formation is selective in ways that public importance does not explain, with African inventory concentrated overwhelmingly in football while salient civic events produce little or no inventory, and Latin American inventory deeper but dominated by Venezuela, where attention to prospective United States military action sustains the largest civic cluster in the data. Legibility orders the inventory steeply, with sports and elections near the top of the scale and conflict at the bottom. In a formation test against an externally assembled frame of 131 civic events, legibility predicts listing in the expected direction but falls short of pre-specified acceptance criteria, while among listed contracts the relation between legibility and trading value is negative, as a model of selective listing implies and as we predicted before estimation. Prediction-market inventories therefore measure what platforms can settle as much as what traders believe, and reading them as maps of public interest conflates the two.

[1] Harvard Business School (aadegbenro@mba2026.hbs.edu). I am grateful to seminar participants and colleagues for helpful comments. All errors are my own.

## 1. Introduction

Prediction markets transform uncertainty into contracts, and the transformation is not automatic. Before a market can aggregate any information, a platform must decide that an event can be described with sufficient precision, resolved through acceptable sources, closed at an administrable time, and defended as a legitimate product. The first empirical margin in prediction-market activity is therefore not price accuracy but market formation.

We study that formation margin for Africa-topic and Latin America-topic prediction markets, asking which regional uncertainties became listed contracts and which did not. The question matters because the observed inventory of prediction markets is commonly read as a map of public interest or trader demand, when it also reflects the institutional grammar of tradability, namely what can be converted into contract form, settled without excessive dispute, and made attractive enough to justify the cost of listing.

A pair of contrasts motivates the inquiry. Between January 2022 and November 2025, global prediction-market platforms listed 292 contracts on the Africa Cup of Nations, a football tournament, and those contracts attracted about $54.4 million in observed trading value. Over the same window, the civil war in Sudan, among the deadliest conflicts of the decade, appears in 4 direct contracts with about $41,000 in observed value, and the 2023 Nigerian presidential election, the largest democratic exercise on the continent that year, appears in 3 candidate contracts with about $4,600.[2] Neither public importance nor public attention explains this ordering. Something about the structure of these uncertainties, rather than their significance, governs which of them become markets.

This paper makes listing itself the dependent variable. Rather than asking whether prediction-market prices were accurate once contracts existed, we ask which Africa-topic and Latin America-topic uncertainties entered the contract inventory, which of those contracts attracted observed trading activity, and what the resulting pattern reveals about the conditions under which uncertainty becomes tradable.

The analysis has three components. The first is an audited dataset. We rebuild the population of Africa-topic and Latin America-topic contracts on Polymarket and Kalshi from source records, validate count recall by capture-recapture, validate value recall by probability-proportional-to-size sampling over a $49.9 billion platform-volume universe, stress-test false-positive collisions, and reconcile every layer of exclusion, arriving at 6,047 contracts carrying $486.6 million in observed value. The data construction is reported in unusual detail because, in this setting, it is part of the identification problem: value is so concentrated that a single missed cluster of contracts can reverse a regional conclusion, and our own validation uncovered and repaired exactly such an error. The second component is a measurement instrument. We reduce the intuition that some uncertainties are easier to settle than others to a three-dimension ordinal codebook, settlement legibility, and score 451 units under a frozen protocol with two independent coders, reaching ordinal reliabilities of 0.92 and 0.96 on the primary dimensions and validating the output against blind human benchmarks at two stages, one of which failed and forced a revision we document. The third component is a pair of pre-registered tests. A formation test asks, on an externally assembled frame of 131 salient civic events, whether legibility predicts which events acquired at least one listed contract, conditional on public attention. A selection test asks, among contracts that did form, how legibility relates to trading value, where a model of selective listing makes a sharp and counterintuitive prediction.

[2] All dollar figures in the paper are observed trade-derived value, defined precisely in Section 2. They measure activity in matched trade records, not platform-reported volume and not order-book liquidity.

Our organizing concept is settlement legibility. An uncertainty is settlement-legible when it can be reduced to a repeatable event template, governed by determinate settlement evidence, and closed at a knowable time. The term adapts Scott's (1998) account of legibility to the platform setting, where the relevant capacity is not the state's ability to render society administrable but the platform's ability to render uncertainty contractible. The construct is deliberately narrow, excluding importance, familiarity, attention, and moral seriousness, which shape formation through other channels but are distinct from whether a market can be worded and settled.

The distinction is useful because public importance alone does not determine formation. African and Latin American public life contains elections, wars, coups, referenda, court cases, leadership transitions, commodity shocks, monetary decisions, diplomatic crises, and sports tournaments. Some of these events have scheduled dates, named actors, official result sources, and repeatable templates. Others are multilingual, source-fragmented, episodic, ethically sensitive, or dependent on contested classifications such as control, invasion, recognition, resignation, or ceasefire. The former are easier to list even when the latter are socially more consequential.

The descriptive findings are sharp. Africa is not absent from prediction markets: Africa-topic contracts span 53 topic countries and generate about $92.8 million in observed value. But 77 percent of that value sits in sports, and the Africa Cup of Nations alone accounts for 58.6 percent, making AFCON the benchmark case of how Africa-topic uncertainty scales when contract form is standardized. Latin America generates more than four times Africa's observed value, yet the contrast is not a general regional demand result. Venezuela alone accounts for 423 contracts and about $213.1 million, or 54.1 percent of Latin America observed value, and the highest-value Venezuela contracts concern prospective United States military action and the survival of the Maduro government. Once Venezuela and external-US target contracts are separated, Latin America's durable civic depth is election inventory across many countries, not security-market depth.

The legibility instrument organizes this composition. Mean legibility on the primary zero-to-four scale is 3.99 for sports and 3.98 for elections, then 2.52 for politics and governance, 1.36 for foreign policy, and 0.67 for security and conflict. The ordering is not imposed by sector labels; it follows from the codebook's dimensions, because sports and elections have repeatable templates and determinate settlement records while conflict and governance require contested classifications. Sports contracts were coded blind alongside everything else, so their position at the ceiling of the scale functions as a manipulation check on the instrument rather than an assumption built into it.

The formation test is disciplined by criteria we fixed before estimation, and the criteria are not met. On the 131-event frame, the formation rate in the top legibility tercile exceeds the bottom tercile by 10.1 percentage points, half the pre-registered margin, and the legibility coefficient in a salience-controlled logit is 0.433 with $p = 0.056$, short of the pre-registered significance bar. The estimate is economically meaningful, roughly 10 percentage points of formation probability per legibility point at the sample means, and every point estimate moves in the predicted direction, but we report the test as failed because promoting a near-miss after the fact is exactly what pre-registration exists to prevent. A power analysis quantifies what the frame could and could not have detected: the design reaches conventional power only for effects roughly one and a half times as large as the one estimated, so the honest summary is that legibility is directionally supported and statistically marginal in a frame powered for large effects only.

The selection test, by contrast, confirms its registered prediction. A simple threshold model of listing, in which attention raises expected fee revenue and legibility lowers settlement cost, implies that conditioning on formation induces negative dependence between legibility and attention among listed contracts even when the two are independent in the population. Among the 269 formed contracts with positive observed value, the legibility coefficient on log value is negative and significant in both ordinary and value-weighted regressions, at −0.78 and −0.49 with country-clustered standard errors. The Venezuela complex illustrates the mechanism: contracts on United States military engagement and on Maduro's hold on office sit at the bottom of the determinacy scale yet form the largest civic value cluster in the data, because the underlying uncertainty commands attention large enough for platforms to accept a heavy settlement burden. We wrote the negative value-weighted prediction down before estimation, and we also disclose the registered prediction we got wrong, a count-basis correlation we expected to be weakly positive.

Taken together, the results support a reading of prediction-market inventories that differs from the common one. An inventory reveals two things at once: the beliefs and activity that arise after listing, and the set of uncertainties that platforms were able and willing to convert into contracts. Because prices are observed only after this institutional screen has operated, studies that condition on listed markets risk treating the screen as neutral when it is not. The data do not show that Africans prefer sports or that Latin Americans prefer politics. They show that global platforms filter uncertainty through event grammar, source hierarchy, settlement burden, and trader attention, and that the filter passes African football easily while African civic uncertainty, thin but rising, remains harder to convert into repeated, trusted inventory.

We also make a methodological contribution that we regard as inseparable from the substantive one. Topic-market datasets require validation beyond keyword extraction, because value is power-law distributed, geography labels collide with team names, acronyms, cities, and cultural references, and a single missed head cluster can change the substantive conclusion. Our validation battery uncovered a recall error in our own extraction pipeline that understated Venezuela, the largest country in the data, by roughly a third; Section 3 documents the discovery, the repair from source records, and the audits that bound what remains. We report this not as a confession but as a design principle: in regional prediction-market research, data construction is part of the empirical argument, and the paper shows what an auditable version of that argument looks like.

Two boundaries on interpretation apply throughout. The unit of analysis is topic geography: a Nigeria-topic market is a market about Nigeria, not evidence that Nigerian users traded it, and a Venezuela-topic market may reflect United States foreign-policy attention rather than Venezuelan demand. And the data measure platform inventory and observed trade-derived activity, not participant location, liquidity origin, or unmet local demand. Section 2 develops these boundaries into a precise statement of the estimand.

The remainder of the paper is organized as follows. Section 1.1 places the contribution in the related literature. Section 2 defines the estimand and the measurement choices. Section 3 documents data construction, the recall error and its repair, and the validation battery. Section 4 reports the descriptive composition of the inventory. Section 5 develops the settlement-legibility instrument and its reliability evidence. Section 6 presents the listing model, the pre-registered formation test, and the selection test. Section 7 discusses implications, Section 8 states limitations, and Section 9 concludes. Appendix E collects proofs, estimation details, complete coefficient tables, and the power analysis.

## 1.1 Related literature

Our work connects the empirical literature on prediction markets, which studies prices on contracts that already exist, with the market-design and economic-sociology traditions that treat the existence of a market as the object to be explained. We discuss each strand and our contribution relative to it.

A first literature studies the informational content of prediction-market prices. Its intellectual foundation is the claim that markets aggregate dispersed knowledge (Hayek 1945), developed for event contracts by Wolfers and Zitzewitz (2004, 2006) and Snowberg, Wolfers, and Zitzewitz (2013), advocated for policy use by Arrow et al. (2008), and supported by long-run forecasting evidence such as Berg, Nelson, and Rietz (2008). This literature conditions, almost by construction, on contracts that platforms chose to list. Our contribution is to study the choice itself: we measure the prior selection margin that determines which uncertainties ever reach the price-formation stage, and we show that this margin is steeply ordered by a property, settlement legibility, that has no necessary relation to the informational value of the underlying uncertainty.

A second literature studies mechanisms and microstructure, including scoring rules and liquidity provision for event contracts (Hanson 2007) and the general microstructure of price formation (Madhavan 2000). Recent empirical work brings this apparatus to Polymarket and Kalshi, documenting calibration, trader skill, and behavioral biases on existing contracts (Reichenbach and Walther 2025; Bürgi, Deng, and Whelan 2025; Le 2026), and new lifecycle datasets record listing and resolution events at scale without analyzing listing as an outcome (Jia et al. 2026; Gebele and Matthes 2026). We complement this work at the extensive margin: where it asks how listed contracts trade and price, we ask which contracts come to exist.

A third tradition, from market design and economic sociology, treats markets as constructed institutions rather than naturally occurring ones. Callon (1998) and Çalışkan and Callon (2009) analyze the calculative and institutional work required to qualify goods as exchangeable; MacKenzie and Millo (2003) show how a derivatives market had to be assembled, legitimated, and made calculable before it could price anything; Roth (2007) documents markets that fail to exist because the transaction is repugnant. We study markets that fail to exist for a different reason, verifiability, and we draw the economic content of that reason from the incomplete-contracts distinction between outcomes that are observable and outcomes that are verifiable by third parties (Grossman and Hart 1986; Hart 2017). Our settlement-determinacy dimension is an operational, codable measure of that distinction, and our results give the constructed-markets perspective a quantitative footing in a setting where the candidate universe of markets can be enumerated.

Our empirical design imports a strategy from the media-economics literature on gatekeeping. Eisensee and Strömberg (2007) construct an external universe of natural disasters and ask which received news coverage; McCarthy, McPhail, and Smith (1996) do the same for demonstrations and media attention. The essential move is that the denominator is assembled from sources independent of the outcome, so that selection can be measured rather than assumed. We apply that selection-on-events design to contract listing, assembling a frame of 131 elections, referenda, coups, conflict onsets, and irregular leadership exits from electoral calendars and conflict datasets, freezing it before matching, and measuring which events formed markets.

Closest to our question are three antecedents. Silber (1981) treats contract innovation on commodity futures exchanges as the outcome to be explained, but without an external universe of candidate contracts; we extend the question from exchange-designed commodity contracts to event contracts, where the candidate universe can be assembled independently. Rhode and Strumpf (2013) document the long international history of political betting markets without modeling formation. And a recent ethnography of Polymarket describes the platform's markets team as an epistemic filter deciding which uncertainties become tradable (Rohanifar, Ahmed, and Sultana 2026); our analysis quantifies the output of precisely that filter. To our knowledge, no prior study measures prediction-market formation quantitatively against an externally assembled event frame, and none examines prediction-market activity on African or Latin American topics systematically. Those are the two gaps we fill.

Finally, our measurement strategy draws on content-analysis methodology and its recent extension to model-assisted annotation. The reliability framework is Krippendorff's (2004); the protocol of frozen codebooks, blinded independent coding, and chance-corrected agreement follows that tradition directly. For execution we use model-assisted classification under human-designed anchors, following evidence that such annotation can match or exceed crowd performance when properly supervised (Gilardi, Alizadeh, and Kubli 2023; Törnberg 2023; Ziems et al. 2024), and we discipline it with blind human benchmarks at two stages, including one failure that forced a codebook revision we report in full.

## 2. Estimand and measurement

The interpretation of the evidence depends on three measurement choices. First, we study contracts rather than traders. The dataset identifies what platforms listed and what traded in the records we observe. It does not identify trader residence, liquidity origin, or local demand. Claims about African or Latin American users would therefore be outside the evidentiary design.

Second, the geography variable is topic geography. A contract about an election in Chile, a military strike against Somalia, or an AFCON match involving Senegal is assigned to the country or countries that the contract is about. This is the right unit for a study of platform inventory, but it is not a proxy for where market participants live. In particular, a United States military-action market about Somalia is a Somalia-topic market, but not an internally generated Somalia civic market.

Third, we distinguish inventory from activity. Contract counts measure the extensive margin of listing. Observed trade-derived value measures activity among the contracts for which matched trade records are available. For Kalshi, observed value is estimated from contract count multiplied by trade price where matched trade records exist. For Polymarket, it is based on the USDC leg of matched trades. This measure is closer to realized trading activity than platform metadata alone, but it is not complete order-book liquidity and does not capture depth, spreads, immediacy, or unobserved trades.

These restrictions are not incidental caveats; they define the estimand. We estimate the regional and sectoral composition of listed topic markets and observed trade-derived activity under a documented source pipeline. We do not estimate regional demand, welfare, forecast accuracy, or the social importance of the underlying events.

## 3. Data construction, validation, and sensitivity design

### 3.1 Source data and unit of analysis

The source material is the prediction-market-analysis archive assembled by Jon Becker, a public structured dump of Polymarket and Kalshi market metadata and matched trade records, in the snapshot collected in early February 2026. The archive contains 408,863 Polymarket market records and several million Kalshi market records, together with the matched trade files from which observed value is computed. Direct calls to the platforms' public APIs are used only as supplementary enrichment and verification, not as the base data. We work from these source records rather than from a curated export because the empirical problem is partly one of recall. When a market appears to be absent, it is necessary to know whether it was absent from the platform, missed by our candidate extraction, or removed by our scope rules, and only access to the full source archive makes those three cases distinguishable. This design choice proved consequential: the validation exercises in Sections 3.2 through 3.6 caught a recall error precisely because the full archive was available to audit our own pipeline against.

The unit of analysis is the listed contract. A contract enters the candidate universe if its title, question, subtitle, slug, ticker, event text, or associated metadata indicates a direct Africa-topic or Latin America-topic reference under the geography codebook. The paper-scope layer then removes false positives, broad or indirect region-only mentions, non-civic recovery artifacts, and categories that do not inform the market-formation question.

The repaired candidate extraction identifies 6,908 contracts as possible Africa-topic or Latin America-topic candidates. The paper-scope review retains 6,047 of these contracts and excludes 861: 537 are false positives, contracts that matched a geography term without being about the region, and 324 are contracts whose regional reference is too broad or indirect to inform the formation question. One further exclusion applies to the main analysis. Kalshi lists a family of recurring daily contracts on which song tops the music chart in Brazil or Costa Rica; there are 888 such contracts. They are genuine country-topic inventory, but they are a mechanical product template with almost no trading, adding only $618.17 in observed value, and including them would let one entertainment product family dominate every count-based comparison while changing no value-based conclusion. We therefore exclude them from the analytical sample and report the paper-scope totals with and without them.[3]

The event-frame window equals the observed-trade window, from the first matched trade on 2022-01-27 to the last matched trade on 2025-11-25, because formation must be measurable for every frame event; events after trade coverage ends would mechanically code as unformed.

| Sample layer | Contracts | Purpose |
|---|---|---|
| Repaired candidate set | 6,908 | Broad extraction net before paper-scope exclusions |
| Paper-scope dataset | 6,047 | Curated Africa-topic and Latin America-topic contracts |

[3]The music-chart family illustrates a general feature of contract counts on these platforms: counts are sensitive to platform product design, because a single template can spawn hundreds of near-identical contracts. This is one reason the paper treats observed value, rather than contract counts, as the more robust margin.

| Sample layer | Contracts | Purpose |
|---|---|---|
| Main analytical sample | 5,159 | Paper-scope dataset excluding 888 low-value music-chart contracts |

### 3.2 The measurement problem

Initial validation uncovered a systematic recall error in our own extraction pipeline, and the discovery process matters for how much confidence the repaired data deserve. Rather than trusting the pipeline's output, we audited it from first principles against the source archive: we searched the raw market records directly for contracts that should have been in our dataset and found markets that were present in the archive but absent from our extraction. Tracing those cases backward through the pipeline identified the mechanism. The geography prefilter, a speed optimization that screens markets before the full geography matcher runs, admitted a market only when its text contained a country alias or demonym. It therefore silently dropped markets that name a country only through its leader, party, or armed actor. Markets about Maduro and Venezuela, Lula and Brazil, Sisi and Egypt, or the RSF and Sudan could be discarded before curation ever saw them.

This is measurement error, not a cosmetic data-cleaning issue. Because observed value is highly concentrated, a small number of missed head markets can change the regional interpretation. The repair therefore returns to the source records, reproduces the failure mechanism, adds only collision-screened high-precision actor terms to the prefilter, rebuilds the candidate set and trade aggregates from the archive, and re-runs every downstream output. The repaired pipeline was then re-audited with the independent checks described in Sections 3.3 through 3.6, which test recall and precision by methods that do not depend on the pipeline being audited. The repair is conservative by design, since broad political terms would raise recall but would also introduce high-value false positives.

The final repair increases the paper-scope dataset by 428 included contracts and $100.5 million in observed value. The dominant correction is Venezuela. Paper-scope Venezuela rises from 329 to 423 contracts and from $135.8 million to $213.1 million in observed value. The final paper-scope bridge is +94 Venezuela contracts and +$77.3 million. The raw candidate Venezuela bridge is 103 contracts and $79.8 million, while the all-Maduro raw bridge is about $80.6 million. The prior Venezuela denominator therefore understated the corrected Venezuela total by about 36 percent.

### 3.3 Capture-recapture on count completeness

The first validation exercise asks whether the keyword system missed a large class of Africa/Latin America references by count. We draw a reproducible uniform random sample of 2,000 Polymarket markets from a 408,863-market universe. Two classification passes are then applied. Classifier A is the production geography keyword automaton run without the prefilter gate. Classifier B is an independent semantic classification pass, followed by researcher review of the leakage cases. The reference population is the set of markets that reference an African or Latin American country directly or through a leader, party, or actor.

The 2 by 2 result was:

| Detector result | Count |
|---|---:|
| Both classifiers | 30 |
| Keyword only | 19 |
| Semantic classifier only | 26 |
| Neither | 1,925 |

Using the Chapman bias-corrected Lincoln-Petersen estimator,

$$\hat{N} = ((n_A + 1)(n_B + 1) / (m + 1)) - 1$$

with $n_A = 49$, $n_B = 56$, and $m = 30$, the estimated count prevalence was 90.9 Africa/Latin America references in the sample, with an approximate 95 percent interval of 77.7 to 104.2. This implies a count prevalence of about 4.5 percent of all Polymarket markets, with a 95 percent interval of 3.9 to 5.2 percent. Appendix E.8 derives the variance and interval.

The point estimate is less important than the leakage review. Of the 26 markets found by the semantic classifier and missed by the keyword pass, 24 are sports or out of scope. Civic in-scope keyword recall is 6 of 7, or 86 percent. The one civic in-scope miss is a Maduro-cluster Venezuela market, which belongs to the already identified prefilter failure class.

This exercise is diagnostic rather than dispositive. The keyword and semantic classifiers are positively correlated, so the capture-recapture independence assumption is not satisfied. The result should therefore be read as evidence about the failure mode, not proof of perfect completeness. Its substantive value is that it recovers the same bug class and does not reveal a second count-recall failure of comparable importance.

### 3.4 Value-weighted PPS audit

The second validation exercise targets the weakness of uniform count sampling. Prediction-market value is extremely concentrated. In the Polymarket value universe used for this audit, trade value has a Gini coefficient of 0.945. A uniform sample can miss rare but substantively decisive head clusters; the Maduro repair is precisely such a case.

We therefore use systematic probability-proportional-to-size sampling over 330,310 Polymarket markets with positive platform volume, covering a platform-volume universe of about $49.9 billion. The audit draws 2,500 selection points and produces 1,884 distinct markets. Each market is evaluated

with a production-faithful keyword classifier, a broader keyword classifier, and an independent semantic classifier, then manually reviewed in the high-value civic slice.

The reviewed high-value civic slice contained $248.9 million in in-scope Africa/Latin America value. The production pipeline captured $156.4 million, or 62.8 percent. The recoverable gap was $92.5 million, or 37.2 percent. That gap consisted of the known bug class, with about $56.6 million in Maduro/Venezuela platform-volume PPS signal and about $35.9 million in Sao Paulo/Brazil signal. No additional high-value civic gap surfaced outside the known prefilter allow-list problem.

| PPS-reviewed high-value civic slice | Value | Share |
|---|---:|---:|
| Total in-scope civic Africa/Latin America value | $248.9m | 100.0% |
| Captured by production | $156.4m | 62.8% |
| Recoverable known-gap class | $92.5m | 37.2% |
| Additional high-value civic gap outside known bug surfaced by independent classification | $0.0m | 0.0% |

This is a value-weighted audit of observed and independently classified markets, not a proof of perfect completeness. It cannot observe markets missed by both the production classifier and the independent semantic classifier. Its contribution is that it targets the part of the distribution where a missing cluster would change the dollar findings. The audit finds the Venezuela and Sao Paulo repair class and does not reveal another high-value civic class.

### 3.5 Collision audit and precision controls

The third validation test asked what would happen if the extraction system simply added broader geography and political terms. The answer is that precision would collapse. The naive full keyword net carried roughly $479 million of false-positive platform-volume value. Examples include San Jose Sharks markets misread as Costa Rica, Crystal Palace markets misread as Morocco, Valencia football markets misread as Venezuela, Panama strings in poker or Federal Reserve contexts, CAR strings in Carolina Panthers markets, and Alexandria Ocasio-Cortez references misread as Egypt.

This collision audit justifies the conservative repair design. The production prefilter caused the Maduro bug because it excluded some political terms, but it also prevented many high-value false positives. The correct repair was not to add all political terms but to add a curated, boundary-aware, collision-screened allow-list and to reject unsafe bare acronyms such as da, pdp, slpp, and npp.

### 3.6 Recovery audit and paper-scope reconciliation

The post-repair pipeline was audited at the level of individual contract records. The candidate extraction gained 572 candidate ids and lost 19 old ids, with a net gain of 553. The 19 removed records were all old South Africa false positives caused by unsafe bare da matching Brazilian football or unrelated title text. No genuine South Africa contract was identified in the removed set. The independent review also found and repaired wrong-country additions involving Sao Paulo FC, Vasco, Philippine PDP, and Sri Lankan SLPP and NPP collisions.

The paper-scope recovery slice contains 282 recovered leader-reference contracts and $93.2 million in observed value, with zero sports and zero entertainment sectors. The candidate-versus-paper-scope Venezuela bridge is also reconciled. The paper-scope layer retains 94 of those contracts, carrying $77.3 million. The $2.48 million difference is intentional, because the paper-scope rules exclude a Maria Corina Machado Nobel Peace Prize market, a Musk x Maduro fight spectacle market, and seven Venezuela award and search-ranking contracts.

The same repair did not produce an Africa correction on the same scale. The repaired paper-scope recovery adds 21 Africa contracts and about $0.603 million in observed value, including 16 recovered leader-reference contracts worth about $0.101 million and 5 Egypt contracts worth about $0.502 million. That is less than 1 percent of Africa observed value and does not alter the Africa interpretation. The material repair is Latin America, dominated by Venezuela.

This reconciliation matters because it prevents the repair from turning into an indiscriminate expansion. The repaired denominator recovers civic and political actor markets, while excluding non-civic spectacle and entertainment contracts.

### 3.7 Robustness and sensitivity tests

The substantive findings are tested under several exclusions and decompositions:

| Test | Why it is used | What it shows |
|---|---|---|
| Music-chart exclusion | Prevents low-value repeated Kalshi music contracts from inflating Latin America counts | LatAm count falls by 888, value falls by under $1,000 |
| AFCON exclusion | Tests whether Africa sports depth is a general sports result or AFCON-concentrated | Africa sports value falls from $71.8m to $17.5m |
| Sports exclusion | Separates civic and economic activity from sports inventory | Africa non-sports is $20.9m; LatAm non-sports is $372.0m |
| Venezuela exclusion | Tests whether LatAm civic depth survives its largest stress case | Security/conflict weakens sharply; elections remain strong |
| External-US target exclusion | Separates target-country geography from internally generated civic-security depth | LatAm and Africa security/conflict values fall sharply |
| Equal-missingness sensitivity | Shows how incomplete trade-derived coverage could scale if missing contracts resembled covered contracts | Useful scale check, not an identified estimate |
| Chi-square and Cramer's V diagnostic | Describes region-sector association without treating contracts as independent social observations | Association remains after music removal, but smaller |

The chi-square diagnostic is descriptive only. In the full paper-scope sample, the sector-region association has chi-square 892.6, 13 degrees of freedom, Cramer's V 0.384, and n = 6,047. After excluding music-chart contracts, the statistic falls to chi-square 424.9, 12 degrees of freedom, Cramer's

V 0.287, and n = 5,159. After excluding music charts and the other category, it is chi-square 359.7, 11 degrees of freedom, Cramer's V 0.278, and n = 4,666. These statistics describe inventory composition. They should not be read as inferential tests on independent social events because contracts are platform-designed units and mechanically repeated market families violate independence.

## 4. Descriptive findings

### 4.1 Regional scale

After the music-chart exclusion, the analytical sample contains 5,159 contracts. Table 1 reports the pooled Polymarket and Kalshi sample. Appendix C reports platform and time-window checks. Because Polymarket accounts for 97.3 percent of observed value while Kalshi accounts for 23.0 percent of contracts, value claims are more platform-robust than count claims.

| Region | Contracts | Topic countries | Contracts with observed value | Share with observed value | Observed value | Observed trades |
|---|---|---|---|---|---|---|
| Africa | 1,768 | 53 | 1,460 | 82.6% | $92.8m | 1,115,247 |
| Latin America | 3,391 | 32 | 2,983 | 88.0% | $393.8m | 3,887,032 |

The table shows two margins. Africa has wider country coverage, with 53 topic countries compared with 32 for Latin America, while Latin America has deeper observed value, with $393.8 million compared with $92.8 million. The comparison is not a demand statement but a statement about platform inventory and observed trade-derived activity around regional topics.

The larger difference is compositional. Africa-topic activity is sports-led, while Latin America-topic activity is non-sports-led, with Venezuela and elections carrying much of the observed value.

### 4.2 Sports versus non-sports

The sports/non-sports split gives the cleanest descriptive contrast. It separates a highly standardized class of contracts from the civic and economic categories that motivate the formation question.

| Region | Category | Contracts | Observed value |
|---|---|---|---|
| Africa | Sports | 1,201 | $71.8m |
| Africa | Non-sports | 567 | $20.9m |
| Latin America | Sports | 1,364 | $21.8m |
| Latin America | Non-sports | 2,027 | $372.0m |

Africa sports value is more than three times Latin America sports value, while Africa non-sports value is about one eighteenth of Latin America non-sports value. This is the central empirical contrast. It does not imply that African users prefer sports; it implies that global prediction-market platforms converted Africa-topic sports into traded markets much more successfully than they converted Africa-topic civic and economic uncertainty.

The pattern is not an artifact of the full time window. In the all-years analytical sample, Africa sports is 77 percent of Africa observed value and non-Venezuela Latin America observed value is 1.9 times Africa observed value. The corresponding figures are 79 percent and 1.9 times for markets created from 2024, and 80 percent and 1.8 times for markets created from 2025. The 777 analytical-sample contracts without creation timestamps do not generate the result, since they carry $12.6 million in Africa observed value and $112.3 million in Latin America observed value. In the dated portion of the sample, African non-sports observed value grows from $0.26 million in 2023 to $7.94 million in 2025. The African civic pattern is therefore thin but rising, not static absence.

### 4.3 Africa and AFCON

The Africa sports result is concentrated in a specific repeated tournament class. AFCON matters analytically because it offers the combination of named teams, scheduled fixtures, official scores, and repeatable contract templates that lowers listing and settlement costs.

| Africa scenario | Contracts | Observed value | Share of Africa observed value |
|---|---:|---:|---:|
| All Africa | 1,768 | $92.8m | 100.0% |
| Africa sports | 1,201 | $71.8m | 77.4% |
| AFCON only | 292 | $54.4m | 58.6% |
| Africa excluding AFCON | 1,476 | $38.4m | 41.4% |

AFCON match markets account for 264 contracts and about $53.1 million. AFCON outright markets account for 28 contracts and about $1.25 million. Together, AFCON match and outright markets account for 75.7 percent of Africa sports observed value.

Once AFCON is removed, Africa's remaining pattern becomes more mixed. Sports remains important, but its share of remaining observed value falls to 45.5 percent. Security/conflict rises to 29.0 percent and elections rise to 13.7 percent. AFCON therefore sharpens the argument, as the clearest Africa-topic case in which repeated event grammar, official sources, and trader familiarity combine to produce scalable market inventory.

### 4.4 Latin America, Venezuela, and elections

Latin America appears civic-heavy, but that conclusion depends on how Venezuela is handled.

| Latin America scenario | Contracts | Observed value | Share of Latin America observed value |
|---|---:|---:|---:|
| All Latin America, excluding music | 3,391 | $393.8m | 100.0% |
| Venezuela only | 423 | $213.1m | 54.1% |

| Latin America scenario | Contracts | Observed value | Share of Latin America observed value |
| --- | --- | --- | --- |
| Latin America excluding Venezuela and music | 2,968 | $180.7m | 45.9% |

With Venezuela included, Latin America observed value is concentrated in security/conflict and elections.

| Sector | Observed value | Share of Latin America observed value |
| --- | --- | --- |
| Security/conflict | $124.3m | 31.6% |
| Elections | $118.9m | 30.2% |
| Other | $87.6m | 22.3% |
| Sports | $21.8m | 5.5% |

After Venezuela is removed, the robust civic story becomes election-led.

| Sector excluding Venezuela | Observed value | Share of non-Venezuela Latin America observed value |
| --- | --- | --- |
| Elections | $113.9m | 63.0% |
| Sports | $21.5m | 11.9% |
| Other | $15.1m | 8.3% |
| Macroeconomy | $10.6m | 5.9% |
| Politics/governance | $6.2m | 3.4% |
| Security/conflict | $4.6m | 2.6% |

The broad Latin America civic claim survives the Venezuela exclusion. The broad Latin America security/conflict claim does not. Outside Venezuela, the durable civic result is election inventory across many countries.

### 4.5 External-US target markets

Some of the highest-value security/conflict markets are not local civic markets in the ordinary sense. They are markets about United States action toward target countries. The target-country tag is valid for topic geography, but it should not be interpreted as local security-market depth.

| External-US military target group | Contracts | Observed value |
| --- | --- | --- |
| All flagged contracts | 198 | $132.3m |
| Venezuela security/conflict | 134 | $117.6m |
| Somalia security/conflict | 18 | $9.1m |
| Mexico security/conflict | 11 | $1.9m |
| Colombia security/conflict | 11 | $1.2m |
| Cuba security/conflict | 11 | $1.2m |

| External-US military target group | Contracts | Observed value |
|---|---|---|
| Nigeria security/conflict | 13 | $1.2m |

Removing external-US target markets sharply changes the apparent security/conflict story.

| Scenario | Regional observed value | Security/conflict value | Security/conflict share |
|---|---|---|---|
| Africa, all contracts | $92.8m | $11.1m | 12.0% |
| Africa excluding AFCON | $38.4m | $11.1m | 29.0% |
| Africa excluding external-US target contracts | $82.4m | $0.8m | 1.0% |
| Africa excluding AFCON and external-US target contracts | $28.1m | $0.8m | 3.0% |
| Latin America, excluding music | $393.8m | $124.3m | 31.6% |
| Latin America excluding external-US target contracts and music | $271.8m | $2.3m | 0.9% |
| Latin America excluding Venezuela and music | $180.7m | $4.6m | 2.6% |
| Latin America excluding Venezuela, external-US target contracts, and music | $176.3m | $0.3m | 0.2% |

This sensitivity is central to interpretation. Security/conflict is not a broad regional depth result but a concentrated attention cluster around external military action and Venezuela.

### 4.6 Country evidence

Country-level evidence shows why the regional comparison cannot be reduced to "Africa sports, Latin America politics."

| Africa country | Contracts | Observed value | Top sector | Interpretation |
|---|---|---|---|---|
| Somalia | 52 | $9.7m | Security/conflict | Mostly external-US strike target markets |
| Senegal | 69 | $9.3m | Sports | AFCON and football shaped |
| Nigeria | 73 | $9.3m | Sports | Football-led; direct 2023 presidential election markets are tiny |
| Egypt | 108 | $8.5m | Sports | Football-led |
| Morocco | 63 | $7.6m | Sports | Football-led |
| Algeria | 34 | $6.3m | Sports | Sports-led by value |
| South Africa | 266 | $6.3m | Sports | Broader inventory, still sports-led |
| Cote d'Ivoire | 60 | $5.3m | Sports | Sports-led with election activity |
| Cameroon | 54 | $4.1m | Sports | Sports-led with some election activity |
| Tunisia | 43 | $3.3m | Sports | Sports-led |
| Mozambique | 40 | $2.1m | Sports | Sports-led |
| Seychelles | 19 | $1.8m | Elections | Rare Africa election-depth case |

Africa's high-value country cases are mostly football-led. Somalia is the largest exception, but that exception is dominated by external-US strike target markets. Seychelles is analytically useful because it shows that Africa election markets can form when a country-specific election template is listed and traded. It is too small to overturn the aggregate pattern, but it prevents the argument from becoming absolute.

| Latin America country | Contracts after music exclusion | Observed value | Top sector | Interpretation |
|---|---:|---:|---|---|
| Venezuela | 423 | $213.1m | Security/conflict | Drives the security result; external-US and Maduro-status markets dominate |
| Chile | 236 | $30.7m | Elections | Robust election template |
| Bolivia | 189 | $28.1m | Elections | Robust election template |
| Argentina | 274 | $27.7m | Elections | Election-led, with some sports and macro |
| Brazil | 681 | $24.2m | Elections | Music charts removed; value remains election-led and mixed |
| Honduras | 83 | $21.1m | Elections | Robust election template |
| Mexico | 528 | $17.7m | Macroeconomy | Mixed macro, sports, and security |
| Ecuador | 70 | $6.1m | Elections | Election-led |
| Suriname | 42 | $4.7m | Politics/governance | Governance-led |
| Colombia | 180 | $4.2m | Mixed | Sports, elections, and security all present |
| Panama | 39 | $2.9m | Foreign policy | Canal sovereignty cluster |
| Costa Rica | 112 | $2.2m | Elections | Music charts removed; value remains election-led |

Venezuela dominates the security/conflict story, but elections remain large across Chile, Bolivia, Honduras, Argentina, Brazil, Ecuador, Costa Rica, and other countries. That is why we treat non-Venezuela Latin America civic depth as real but election-led.

## 5. Measuring settlement legibility

Sections 3 and 4 establish the inventory pattern. The next question is whether the pattern can be organized by a measurable property of contract design rather than by an ex post description of sectors. We therefore convert settlement legibility into an ordinal instrument. The purpose is not to claim that legibility is the only determinant of market formation but to isolate one platform-side condition that must be satisfied before attention can become a listed contract.

The measurement exercise has three steps. First, we reduce the concept to explicit codebook anchors. Second, we implement a blinded coding protocol and report reliability, including the benchmark that failed. Third, we use the resulting scores to describe the contract inventory and to construct the formation test in Section 6.

### 5.1 From concept to codebook

Settlement legibility is defined at the level of contract form. An uncertainty is easier to list when its event class recurs in a recognizable template, when settlement evidence is determinate, and when the resolution moment can be specified. These properties are conceptually related but empirically separable. We therefore score each unit on three ordinal dimensions, each taking values 0, 1, or 2 from the contract question or event description and its stated resolution source.

**D1 (Template repeatability).** Does the event class recur as something a platform can list repeatedly? A league fixture or a constitutionally scheduled election scores 2. An event class that recurs irregularly, such as a coup, an indictment, or a ceasefire attempt, scores 1. A one-off event scores 0. The codebook treats a single underlying event re-listed under successive deadlines as one-off, since rolling re-listings of one uncertainty are a pricing convenience rather than a repeatable template.

**D2 (Settlement determinacy).** What kind of evidence resolves the contract? A certified ballot, a final score, or a recorded legislative vote scores 2. A reported act that requires classification, such as whether two states "normalize relations" or whether a leader "resigns," scores 1. A contestable judgment, such as whether a force "controls" a city or whether a "military engagement" occurred, scores 0.

**D3 (Closure precision).** Could a trader mark the resolution moment on a calendar in advance? D3 is scored from a decision tree that distinguishes scheduled moments, bounded windows, and open-ended triggers, with an explicit branch for institutions whose calendars exist on paper but not in practice.

Our primary legibility score is the sum of the first two dimensions,

$$L_i = D1_i + D2_i, \; L_i \in \{0, 1, 2, 3, 4\}. \quad (1)$$

Throughout the empirical sections, settlement legibility means exactly what the codebook scores and nothing more. The term itself borrows deliberately from Scott (1998), for whom legibility is the state's capacity to render society readable in standardized, administrable form; here it is the platform's capacity to render uncertainty readable in contract form. The construct deliberately excludes attention, importance, and familiarity, which enter the analysis as a separate salience control rather than as components of the score.

We keep D3 out of the primary score for a reason the validation history makes plain. In the first blind human benchmark, D3 failed its reliability bar because two readings of the original anchor text were each internally consistent, one keyed to dates stated in the question and one keyed to institutional calendars.[4] We rewrote the anchor as the decision tree above and re-benchmarked it on a fresh sample,

[4] The two readings disagree most visibly on questions of the form "who will win the next election in country c?" where no date appears in the text. Under a date-in-text reading the question is open-ended; under an institutional reading it resolves at a scheduled moment whenever country c has a constitutionally fixed cycle. The revised tree adopts the institutional reading and adds the transitional-regime branch, under which an announced date issued by an unelected interim government does not count as a reliable schedule.

where it passed. The conservative choice is to treat the twice-benchmarked two-dimension score as primary and the three-dimension sum as a secondary specification. We report both in Section 6.

To make the anchors concrete, consider how they apply to three contract types that anchor the top, middle, and bottom of the scale. A scheduled AFCON fixture, "Will Senegal beat Cameroon in their group-stage match?", scores D1 = 2 because tournament fixtures are a template the platform lists repeatedly, D2 = 2 because an official final score settles the contract, and D3 = 2 because kickoff is on the calendar; its primary score is the maximum, L = 4. A presidential-election contract, "Who will win Chile's presidential election?", scores D1 = 2 and D2 = 2 for the same structural reasons, scheduled recurrence and a certified official result. A contract of the form "Will the United States engage militarily with Venezuela by the deadline?" sits at the other end: the event class does not recur as a listable template, so D1 is at most 1; settlement turns on whether some action counts as a "military engagement," a contestable classification, so D2 = 0; and resolution happens only if and when something occurs, so D3 reflects the stated deadline rather than a schedule. Its primary score is L of 0 or 1. The gradient documented below is this logic applied 451 times under blind conditions, not a judgment about which events matter.

### 5.2 Coding protocol and reliability

The scientific content of the instrument lies in the codebook, and the codebook is authored, frozen, and human-owned. We wrote the construct definition, the anchors for each dimension, and the decision rules before any unit was coded, and froze them; every coding decision in the paper traces to that document or to a written ruling extending it, logged rule by rule in the replication materials. Execution of the 451-unit production pass then used model-assisted classification as the measurement apparatus, following recent methodological work showing that supervised annotation under explicit human-designed anchors can match or exceed crowd performance (Gilardi, Alizadeh, and Kubli 2023; Törnberg 2023; Ziems et al. 2024). Two coders worked independently on identical material presented in independently shuffled order, and saw only the contract question and its stated resolution source. We withheld sector labels, trading values, and salience information, so that a coder could not learn that a question was "sports" or "high volume" except from the question text itself. Disagreements were resolved by a third adjudication pass bound to the same codebook, and residual cases that the codebook genuinely underdetermines were settled by written reviewer rulings. The apparatus was then disciplined by the part of the protocol that only a human can supply: blind human benchmarks at two stages, designed in advance and allowed to fail. We report all benchmark scores below and in Appendix D, including the one that failed.

The protocol, in sequence, was the following.

| Stage | What was done | Output |
|---|---|---|
| 1. Construct and codebook | Anchors and decision rules written and frozen before coding | Frozen codebook |
| 2. Blind double coding | Two independent coders, different model families, blinded shuffled inputs | 451 units coded twice |
| 3. Adjudication | Third pass bound to the codebook; underdetermined cases settled by written rulings | Final codes |
| 4. Human benchmark, stage one | 48-unit blind human gold standard against the coded output | D1 and D2 pass; D3 fails |
| 5. Anchor revision | D3 anchor rewritten as an explicit decision tree | Revised codebook |
| 6. Human benchmark, stage two | Fresh 24-unit blind human revalidation of D3 | D3 passes |

We measure reliability with Krippendorff's alpha for ordinal data,

$$\alpha = 1 - D_o \,/\, D_e, \quad (2)$$

where D_o is the observed mean squared ordinal distance between paired codes and D_e is the distance expected under chance assignment from the empirical marginal distribution (Krippendorff 2004). The conventional floor for drawing tentative conclusions is $\alpha \geq 0.67$; we adopted this floor in advance for every reliability claim we make. Appendix E.9 derives the coefficient from first principles and proves the properties, perfect-agreement and chance-correction, that justify its use.

The coded sample contains 451 units. These are 320 contracts, combining all 80 sports contracts from the audit sample with a civic sample built as a census of every civic contract above $1 million in observed value plus stratified random fill, together with the 131 events of the formation frame constructed in Section 6.[5] Inter-coder agreement on this sample is $\alpha = 0.921$ for D1, $\alpha = 0.958$ for D2, and $\alpha = 0.760$ for D3, each above the floor. The blind human benchmark agrees with the final coded output at $\alpha = 0.968$ and $\alpha = 0.918$ on the two primary dimensions. For D3 the first benchmark produced $\alpha = 0.417$, the failure discussed above, and the post-revision benchmark produced $\alpha = 0.772$. We report the failed benchmark rather than discard it because it is informative about the construct. Closure was the hardest dimension to specify precisely, and the validation process is what surfaced that fact.

Two further checks discipline the instrument. Sports contracts, which our conceptual discussion treats as the canonical legible class, were coded blind alongside everything else and score a mean of 3.99 on the 0-to-4 scale, essentially at the ceiling. This functions as a manipulation check. If the anchors did not recover sports at the top of the scale, the instrument would be measuring something other than the concept. The blind human benchmark, in turn, did the work it was designed into the protocol to do: beyond forcing the D3 revision, it surfaced four codes left over from an earlier anchor version, which we corrected before any analysis. An instrument that cannot fail is not a measurement; this one visibly could, twice, and both failures are documented.

[5] The census-plus-fill design reflects the value concentration documented in Section 3. Because observed value is power-law distributed, a simple random sample of contracts would characterize the long tail while missing the head that carries most of the value. The census component guarantees that every economically consequential civic contract is scored.

### 5.3 The legibility gradient in the inventory

Mean primary legibility in the coded inventory is 3.99 for sports and 3.98 for elections, then 2.52 for politics and governance, 1.36 for foreign policy and diplomacy, and 0.67 for security and conflict, a spread of more than three points on a four-point scale. The ordering is not imposed by the sector labels. It follows from the codebook dimensions, since sports and elections have repeatable templates and determinate settlement records while conflict and governance often require contested classifications.

This result formalizes the descriptive contrast in Section 4. Football is not simply popular; it is contractible in a standardized way, with named teams, scheduled fixtures, official scores, and a template that repeats across matches and tournaments. Elections share many of these properties through ballots, dates, candidates, certification procedures, and official sources. By contrast, uncertainty about invasion, control, resignation, recognition, or ceasefire often requires judgment over facts whose meaning is disputed.

The construct is intentionally incomplete, measuring settlement burden rather than attention. Media-mention contracts are mechanically settleable yet small. The Venezuela complex scores near the bottom of the scale while carrying $213.1 million in observed value. Those cases do not invalidate the concept; they identify the selection problem. Low-legibility events can form when attention is sufficiently large. Section 6 tests this implication directly.

## 6. A formation test against an external event frame

The contract inventory alone cannot identify formation because it contains only successful listings. A test of formation requires a denominator of events that could have become markets but did not necessarily do so. We therefore construct an external frame of salient civic events in Africa and Latin America, freeze the selection rules before matching, score the events using the same legibility instrument, and then ask which events acquired at least one paper-scope market.

The test is deliberately conservative. It treats legibility as one determinant of formation and controls for attention, the leading alternative explanation. It also fixes the acceptance rule before estimation, so that a directionally favorable but statistically marginal result cannot be promoted after the fact.

Our hypothesis is the following.

**Hypothesis (Legibility and formation).** *Among salient African and Latin American civic events, the probability that an event acquires at least one listed contract is increasing in settlement legibility, conditional on public attention to the event.*

The conditioning clause matters. Salience is the leading alternative explanation for formation, since platforms may simply list whatever the world is looking at. The test below is a test of legibility conditional on attention, not of legibility alone.

### 6.1 The event frame

The frame contains 131 events inside the study window from January 27, 2022 to November 25, 2025. It is built from sources external to the platforms and frozen, with written selection rules, before matching to the contract data.

Elections and referenda enter from the IFES ElectionGuide calendar. Every national election and national referendum in the 85-country universe with an event day inside the window is included, aggregated to the country-day occasion, giving 97 election occasions and 17 referenda. Coups and attempted coups enter from coup-event datasets and were independently re-verified against contemporaneous reporting, with announced plots that produced no action event excluded, giving 12 events. Armed-conflict onsets enter under a UCDP episode-start rule, giving 3 events. Irregular exits of chief executives, through impeachment, forced resignation, or flight, give 2 events.

The window is applied symmetrically. Three near misses show that the boundary is binding rather than decorative. The January 2022 Burkina Faso coup precedes the window by three days and is excluded. The UCDP episode start for the M23 resumption in the eastern Democratic Republic of the Congo is December 26, 2021, one month before the window opens, and is excluded. The Guinea-Bissau coup of November 26, 2025 misses the closing boundary by one day and is excluded. Each exclusion would have added a salient low-legibility event. Each is excluded because the rule was fixed first.

Every event was scored on the Section 5 codebook under the same blind protocol, and every event carries a salience control,

$$S_e = \max \text{ over } t \in [t_e - 14, t_e + 14] \text{ of daily English-Wikipedia pageviews of the country page, } (3)$$

where $t_e$ is the event day.[6]

A simple listing model rationalizes the hypothesis and yields a second testable implication. A platform lists a contract on event e when expected fee revenue, which rises with attention, covers settlement cost, which falls with legibility, net of a fixed listing cost $\kappa$ and an idiosyncratic shock $u_e$ capturing legal exposure, staffing, and taste,

$$F_e = 1\{ R(S_e) - c(L_e) + u_e \geq \kappa \},\ R' > 0,\ c' < 0.\ (4)$$

[6]Country-page views are a deliberately coarse proxy. Event-specific pages do not exist uniformly across the frame, and their creation is itself an outcome of attention, whereas every country page exists throughout the window. The 14-day symmetric window captures both anticipatory attention to scheduled events and reactive attention to unscheduled ones. Appendix E.10 records the design properties of the proxy, all fixed before any market data were matched.

**Proposition 1.** *Under (4), the listing probability* $Pr(F_e = 1 \mid L, S)$ *is increasing in both legibility* $L_e$ *and salience* $S_e$.

*Proof.* Conditional on $(L_e, S_e) = (l, s)$, listing occurs exactly when $u_e \geq \kappa - R(s) + c(l)$, so $Pr(F_e = 1 \mid l, s) = 1 - G(\kappa - R(s) + c(l))$, where G is the distribution function of the shock. The argument of G is decreasing in s because R is increasing, and decreasing in l because c is decreasing; since $1 - G$ is decreasing in its argument, the listing probability is increasing in both. ■

The hypothesis above is the empirical counterpart of the first comparative static, and the logit in Section 6.2 is its test; Appendix E.1 derives that logit, its likelihood, and its standard errors from (4) under a logistic shock. The model's second implication concerns what survives the filter.

**Corollary 1.** *Conditional on* $F_e = 1$*, legibility and salience are negatively associated even when independent in the population. An illegible event clears the listing threshold only when its salience is extreme.*

Corollary 1 implies that regressions of trading activity on legibility among listed contracts are uninformative about the formation mechanism and should, if anything, run negative. The argument is short. Conditional on listing, the joint distribution of legibility and salience is the population distribution tilted by the survival probability of the threshold event; when the shock density is log-concave, that tilt is log-submodular in the two arguments, so among listed events a higher salience draw makes low legibility relatively more likely, and conversely. Appendix E.2 gives the full proof under a log-concavity condition on the shock distribution that the logistic specification in Section 6.2 satisfies. We test the proposition in Sections 6.1 through 6.3 and the corollary in Section 6.4, and we record here that the corollary's prediction was written down before estimation.

Matching is rule-based, with the full procedure and audit trail retained in the replication materials, including 2,312 rejected candidate pairs. The key rule treats wars as episodes rather than moments, so conflict-onset events match contracts about the conflict episode while point events such as coups match only contracts that reference the occasion itself. This rule was corrected during audit in the direction adverse to the hypothesis. The correction flipped the largest event in the lowest-legibility class, the Sudan civil war, from unformed to formed. Under the final rules, 52 of the 131 events formed at least one contract, through 493 event-contract pairs.

### 6.2 Pre-specified acceptance criteria

A test with a small event frame and a new construct is vulnerable to ex post interpretation. We therefore wrote the acceptance criteria down before estimation. We would treat the hypothesis as supported only if both of the following conditions held.

C1. The formation rate in the top tercile of legibility exceeds the rate in the bottom tercile by at least 20 percentage points.

C2. The legibility coefficient is positive at $p < 0.05$ in the logit

$$\Pr(F_e = 1) = \Lambda( \beta_0 + \beta_1 L_e + \beta_2 \ln S_e + \beta_3 \text{LatAm}_e ), \quad (5)$$

where F_e indicates that event e matched at least one contract, L_e is the two-dimension legibility score, S_e is the salience proxy in (3), LatAm_e indicates a Latin America event, and Λ is the logistic function. We pre-specified a Fisher exact test on a high-low legibility split as the fallback under separation or small-cell problems, and we pre-committed to reporting the two-dimension and three-dimension scores, a run excluding the single borderline frame event, and an elections-and-referenda-only run as sensitivity analyses.

### 6.3 Results

The pre-specified criteria are not met. Formation rates by legibility level are 45.4 percent for events scoring $L = 4$, which are scheduled elections, 11.8 percent for events scoring $L = 3$, which are referenda, and 35.3 percent for events scoring $L = 1$, which are coups, conflict onsets, and irregular exits. The top tercile exceeds the bottom by 10.1 percentage points, half the margin required by C1. In the logit (5), the legibility coefficient is $\beta_1 = 0.433$ with $p = 0.056$, positive but above the threshold required by C2. The salience control is itself near conventional significance ($\beta_2 = 0.485$, $p = 0.050$). The Fisher fallback gives $p = 0.598$. Appendix E.3 reports the complete coefficient tables with confidence intervals, Appendix E.4 the marginal effects, and Appendix E.6 the tabulated sensitivity runs.

The aggregate rates are shown two ways: first by the primary legibility score used for C1, and then by event class.

| Primary legibility score | Events | Formed events | Formation rate | Matched observed value |
|---|---:|---:|---:|---:|
| L = 1 | 17 | 6 | 35.3% | $0.3m |
| L = 3 | 17 | 2 | 11.8% | $0.012m |
| L = 4 | 97 | 44 | 45.4% | $104.5m |

| Event class | Events | Formed events | Formation rate | Matched observed value |
|---|---:|---:|---:|---:|
| Conflict onset | 3 | 2 | 66.7% | $160,729 |
| Coup attempt | 12 | 2 | 16.7% | $92,291 |
| Irregular leadership exit | 2 | 2 | 100.0% | $6,473 |
| National executive or other election | 7 | 3 | 42.9% | $268,615 |
| National general election | 29 | 19 | 65.5% | $56.0m |
| National legislative election | 34 | 6 | 17.6% | $16.1m |
| National presidential election | 27 | 16 | 59.3% | $32.2m |

| Event class | Events | Formed events | Formation rate | Matched observed value |
|---|---|---|---|---|
| National referendum | 17 | 2 | 11.8% | $11,753 |

Excluding the one borderline frame event, the January 2023 Brasília attack, does not change the verdict, and neither does the three-dimension secondary score. Restricting the frame to elections and referenda alone produces a passing result. We report that run as a sensitivity and do not promote it, because selecting the passing subsample after seeing results is exactly what the pre-specification rules out.

The result is therefore disciplined but not null in the substantive sense. The coefficient is positive, its magnitude is economically meaningful, and the point estimates move in the direction predicted by the construct.[7] The correct interpretation is that legibility is directionally supported and statistically marginal in a 131-event frame under criteria fixed in advance. The non-monotonicity is also informative. Referenda are highly settleable and rarely formed, while impeachments and coups are poorly settleable and formed at moderate rates, including both irregular-exit events in the frame. Attention competes with legibility in formation, and in this sample it is strong enough to prevent the primary test from passing. A design limitation follows directly, in that event-level legibility varies mainly across event classes rather than within them, so regression (5) compares classes rather than events within a class.

### 6.4 Legibility and value among formed contracts

A companion test asks whether legibility predicts observed value conditional on formation. The listing model in (4) implies that this is not where a positive formation mechanism should necessarily appear, because formation is a filter that an event can pass either by being easy to settle or by being sufficiently salient to justify a high settlement burden. Conditioning on $F_e = 1$ therefore induces negative dependence between legibility and salience among listed contracts even when the two are independent in the population (Heckman 1979; Elwert and Winship 2014). Since salience drives trading activity, the value relation among formed contracts should be flat or negative, with the strongest negative pull in value-weighted form where the Venezuela complex dominates. We recorded this prediction before estimation.

The data are consistent with that selection logic. Across the 320 coded contracts, the Spearman rank correlation between legibility and observed value is −0.247 ($p < 0.001$). The regression

$$\ln V_i = \gamma_0 + \gamma_1 L_i + \text{region and sector fixed effects} + \varepsilon_i, \quad (6)$$

[7] At the sample means, a one-point increase in L raises the predicted formation probability by roughly 10 percentage points, an effect of the same order as the raw tercile gap. Appendix E.4 reports the exact marginal effects, 0.103 at the covariate means and 0.092 averaged over the frame.

estimated on the 269 contracts with positive observed value, with standard errors clustered on 56 topic countries, gives $\gamma_1 = -0.78$ ($p = 0.027$). The value-weighted variant gives $\gamma_1 = -0.49$ ($p = 0.003$). One record-keeping point is important. The negative value-weighted result was predicted in advance, while the rank-order result was expected to be weakly positive and was not. The selection logic in (4) accounts for both signs, but only one of the two was anticipated.

| **Test A value regression** | **OLS, positive notional** | **Value-weighted OLS, positive notional** |
|---|---:|---:|
| Composite legibility coefficient | -0.782 | -0.489 |
| Clustered standard error | 0.354 | 0.165 |
| p-value | 0.027 | 0.003 |
| Observations | 269 | 269 |
| Topic-country clusters | 56 | 56 |
| Region and sector controls | Yes | Yes |
| Dependent variable | ln observed value | ln observed value |
| Weight | None | Observed value |

Appendix E.5 reports the complete coefficient tables, including the region and sector controls, with clustered standard errors and confidence intervals for every term.

The Maduro cluster illustrates the mechanism. Contracts on United States military engagement with Venezuela and on Maduro's hold on office sit at the bottom of the determinacy scale, yet they are the largest civic value cluster in the dataset. They exist at scale because the underlying uncertainty commands attention large enough for platforms to accept a heavy settlement burden. Conditional on formation, their volume is therefore informative about attention rather than ease of settlement.

### 6.5 Salient African civic events, revisited

The motivating case comparisons now sit inside a measured frame. Direct Sudan civil-war and ceasefire contracts number 4, with about $41,000 in observed value, against 292 AFCON contracts carrying $54.4 million. The 2023 Nigerian presidential election produced 3 candidate contracts and about $4,600. Within the event frame, the Sudan war is one of only 6 formed events among the 17 lowest-legibility occasions, and its observed value remains three orders of magnitude below the football benchmark.

The disciplined conclusion is therefore narrower than the motivating contrast but stronger than a purely descriptive claim. Socially important African civic uncertainty did not convert into traded inventory at anything near the rate of templated sport. The measured legibility gradient across sectors is steep, and the formation test points in the predicted direction but does not pass the evidentiary bar fixed in advance. Settlement legibility therefore earns its role as a measured organizing variable; it is not established here as a standalone causal mechanism.

## 7. Discussion

The results change the interpretation of prediction-market data. Because prices are observed only after a prior institutional screen has operated, a platform inventory reveals two things at once, the beliefs and activity that arise after listing and the set of uncertainties that platforms were able and willing to convert into contracts. Studies that condition only on listed markets risk treating this formation screen as if it were neutral.

Africa's presence in the dataset is substantial, spanning 53 topic countries and $92.8 million in observed value, but the highest-activity Africa-topic inventory is concentrated where contract form is easiest to standardize, in national-team football, tournament fixtures, official scores, and repeated match templates. AFCON is not an exception to the argument. It is the benchmark case showing how Africa-topic uncertainty scales when settlement grammar is strong.

Latin America demonstrates the other side of the same mechanism. Civic and political markets can scale in the same platform environment, but broad regional interpretations remain fragile without decomposition. Venezuela materially changes the regional denominator and drives the security/conflict result. After separating Venezuela and external-US target markets, the durable Latin America civic pattern is election inventory, not generalized security-market depth.

The substantive conclusion is therefore precise. Africa is not absent; it is selectively visible. Latin America does not prove local civic demand; it shows that Latin America-topic civic and election uncertainty became more repeatable traded inventory in this dataset. The difference is a formation result, not a revealed-preference result.

The practical implication is that civic prediction markets require resolution infrastructure as much as question selection. Important events are not automatically listable events. Platforms need source hierarchies, local institutional knowledge, language capacity, dispute rules, harm review, and credible settlement governance. In the absence of that infrastructure, inventory will tilt toward uncertainties that are easiest to specify and settle.

The methodological implication is equally central. Regional prediction-market studies should validate recall and precision at the source-record level, audit high-value head clusters separately from the long tail, distinguish topic geography from participant geography, and decompose repeated templates before making regional claims. In this setting, data construction is not preliminary clerical work. It is part of the identification problem.

## 8. Limitations

Six limitations delimit the interpretation. First, we do not observe trader location. Topic geography cannot be interpreted as participant geography or local demand.

Second, observed trade-derived value is incomplete. It is based on collected trade records where available, not complete order-book liquidity. Missing trade-derived value is uneven by platform and region.

Third, the sample is time-bound but not governed by a single timestamp. Market creation dates in the working sample run from April 2021 to February 2026. Source fetches run from November 2025 to February 2026. Live-update fields extend to April 2026. Kalshi matched-trade timestamp fields run through November 2025. Polymarket contributes most observed value, but the export does not preserve first and last trade timestamps for Polymarket matched trades.

Fourth, the classification system is audited but not omniscient. Capture-recapture and PPS audits provide evidence against a second high-value civic recall failure, but they do not prove that markets missed by both independent classification passes have zero value. Sector labels also simplify ambiguous contracts, especially governance, foreign policy, security, and political media markets.

Fifth, the formation test is intentionally bounded. The legibility instrument compares event classes more than events within a class, and the 131-event frame is powered to detect large effects only. The power analysis in Appendix E.7 quantifies this: the tercile comparison reaches 80 percent power only at a true gap of about 35 percentage points, well above the 20-point bar, and the logit reaches it only at a legibility coefficient roughly one and a half times the estimate, so the design could not reliably have detected an effect of the size we observe. The pre-specified formation criteria were not met. We accordingly treat settlement legibility as a measured organizing variable rather than a demonstrated standalone cause.

Finally, we do not study forecast accuracy. Accuracy and truth timing require a different denominator, reconstructed probability histories, outcome labels, and timing rules. Those analyses should remain separate from the market-formation argument.

## 9. Conclusion

Prediction markets reveal platform formation before they reveal trader beliefs. We study that formation layer for Africa-topic and Latin America-topic contracts on Polymarket and Kalshi. The audited sample shows that Africa is present but selectively visible, with observed value dominated by football and especially by AFCON, while Latin America has deeper non-sports and civic activity that depends heavily on Venezuela for its security and conflict component and on elections everywhere else.

The analysis also carries a methodological lesson. Our own validation uncovered and repaired a recall error that materially understated the largest country in the data, and without source-level rebuilding and independent audit, the regional comparison would have rested on an unstable denominator. Any study that infers regional activity from contract text inherits that risk, because value is concentrated enough that one missed cluster can reverse a conclusion.

Our final claim is narrower and more defensible than a demand story. The data do not show that Africans prefer sports, that Latin Americans prefer politics, or that prediction markets measure local public interest. They show that global platforms filter uncertainty through event grammar, source hierarchy, settlement burden, platform supply, and trader attention. In that filter, African football becomes highly tradable. African civic uncertainty is thin but rising, and it remains harder to convert into repeated, trusted, high-activity market inventory.

The natural next step is to extend the formation denominator. A larger external event frame across more platforms and a longer window would separate within-class from across-class variation and test whether the marginal formation effect of settlement legibility survives beyond the large event-class contrasts documented here.

# Appendix A. Replication and audit trail

The replication package contains the source-data description, codebook, candidate extraction rules, paper-scope inclusion and exclusion rules, audit outputs, and executable scripts needed to rebuild the study tables from the source market and trade records. Machine-specific paths, local command histories, and local environment details are not part of the public manuscript.

The public replication package contains four classes of material:

| Material | Purpose |
|---|---|
| Source-data manifest | Identifies source tables, collection windows, platform coverage, and trade-record availability |
| Candidate and paper-scope codebook | Documents geography terms, boundary rules, sector definitions, inclusion rules, and exclusion rules |
| Validation outputs | Reports capture-recapture, PPS sampling, leakage review, collision audit, recovery audit, and reconciliation results |
| Rebuild scripts and tests | Regenerate the analytical tables and verify the boundary/collision rules |

# Appendix B. Data definitions

| Term | Definition |
|---|---|
| Contract | A listed market contract on Polymarket or Kalshi |
| Topic geography | The country or region that the contract is about |
| Participant geography | The location of traders or liquidity providers, not observed here |
| Observed value | Trade-derived notional from collected matched trade records where available |
| Platform volume | Platform-reported volume metadata, used only in validation where appropriate |
| Paper-scope dataset | The curated set of direct Africa-topic and Latin America-topic contracts after false-positive and scope exclusions |
| Analytical sample | Paper-scope dataset excluding low-value music-chart contracts |
| Settlement legibility | The degree to which an uncertainty can be worded, sourced, observed, closed, and resolved credibly |

# Appendix C. Robustness diagnostics

## Coverage sensitivity

Some contracts do not have observed trade-derived value. The equal-missingness sensitivity below assumes missing contracts have the same mean observed value as covered contracts in the same group. It is a scale check, not an identified estimate.

| Basis | Group | Contracts | Share with observed value | Observed value | Equal-missingness sensitivity |
|---|---|---|---|---|---|
| Paper-scope sample, including music | Africa | 1,768 | 82.6% | $92.8m | $112.3m |
| Paper-scope sample, including music | Latin America | 4,279 | 69.8% | $393.8m | $564.0m |
| Paper-scope sample, including music | Latin America Kalshi | 1,839 | 40.1% | $9.9m | $24.8m |
| Analytical sample, excluding music | Africa | 1,768 | 82.6% | $92.8m | $112.3m |
| Analytical sample, excluding music | Latin America | 3,391 | 88.0% | $393.8m | $447.7m |
| Analytical sample, excluding music | Latin America Kalshi | 951 | 77.0% | $9.9m | $12.9m |

Section 4.1 reports the exact analytical-sample coverage count: 2,983 Latin America contracts with observed value, or 88.0 percent after rounding.

### Platform robustness

The analytical sample is value-dominated by Polymarket but count-mixed across platforms.

| Platform | Region | Contracts | Observed value | Sports share of observed value | Non-sports observed value |
|---|---|---|---|---|---|
| Polymarket | Africa | 1,532 | $89.7m | 77.1% | $20.5m |
| Polymarket | Latin America | 2,440 | $383.9m | 4.6% | $366.2m |
| Kalshi | Africa | 236 | $3.1m | 87.8% | $0.4m |
| Kalshi | Latin America, excluding music | 951 | $9.9m | 41.9% | $5.8m |

Within Polymarket alone, Africa is 77 percent sports by observed value and Latin America non-sports observed value is 17.8 times Africa non-sports observed value. Venezuela accounts for 55.3 percent of Polymarket Latin America observed value. Polymarket accounts for 97.3 percent of analytical-sample observed value. Kalshi accounts for 23.0 percent of analytical-sample contracts but only 2.7 percent of observed value; all excluded music-chart contracts are Kalshi.

### Time-window robustness

The time-window cut addresses whether the composition contrast is an artifact of platform growth or event timing.

| Window | Contracts used | Africa sports share of Africa observed value | Non-Venezuela Latin America observed value / Africa observed value | Venezuela share of Latin America observed value |
|---|---|---|---|---|
| All years, analytical sample | 5,159 | 77.4% | 1.9x | 54.1% |
| Created from 2024 | 4,253 | 79.1% | 1.9x | 45.8% |
| Created from 2025 | 3,894 | 80.4% | 1.8x | 47.4% |

Contracts without creation timestamps are excluded from the dated windows but included in the all-years row. The analytical sample has 777 contracts without creation timestamps: 168 Africa contracts with $12.6 million in observed value and 609 Latin America contracts with $112.3 million in observed value. The undated rows do not hide an Africa non-sports mass; Africa undated non-sports value is $4.0 million, while Latin America undated non-sports value is $108.7 million.

### Chi-square diagnostic

| Basis | Chi-square | df | Cramer's V | n |
|---|---|---|---|---|
| Paper-scope sample | 892.6 | 13 | 0.384 | 6,047 |
| Excluding music-chart contracts | 424.9 | 12 | 0.287 | 5,159 |
| Excluding music-chart contracts and other | 359.7 | 11 | 0.278 | 4,666 |

These diagnostics describe composition. They are not inferential tests on independent social observations.

## Appendix D. Settlement-legibility coding, formation frame, and model details

### D.1 Codebook anchors

The primary score is D1 + D2 on a 0-to-4 scale. D3 is retained as a secondary specification only after the fresh human revalidation passed.

**D1 (Template repeatability).** Does the event class recur as something a platform can list repeatedly? A league fixture or a constitutionally scheduled election scores 2. An event class that recurs irregularly, such as a coup, an indictment, or a ceasefire attempt, scores 1. A one-off event scores 0. The codebook treats a single underlying event re-listed under successive deadlines as one-off, since rolling re-listings of one uncertainty are a pricing convenience rather than a repeatable template.

**D2 (Settlement determinacy).** What kind of evidence resolves the contract? A certified ballot, a final score, or a recorded legislative vote scores 2. A reported act that requires classification, such as whether two states "normalize relations" or whether a leader "resigns," scores 1. A contestable judgment, such as whether a force "controls" a city or whether a "military engagement" occurred, scores 0.

**D3 (Closure precision).** Could a trader mark the resolution moment on a calendar in advance? D3 is scored from a decision tree that distinguishes scheduled moments, bounded windows, and open-ended triggers, with an explicit branch for institutions whose calendars exist on paper but not in practice.

The D3 decision tree asks whether the market resolves at a scheduled moment, such as a match kickoff, election day, or official statistic release. If the institution's schedule is reliable, D3 = 2. If the event is scheduled in form but chronically delayed or transitional, D3 = 0. If the question instead states a deadline or window, such as "by date," "before date," "in month," or "in year," D3 = 1. If the question resolves only if or when something happens, with no schedule and no deadline, D3 = 0.

### D.2 Coding protocol and reliability

The production coding covered 451 units: 320 contracts and 131 external-frame events. Two AI coders from different model families coded independently on blinded, shuffled inputs containing only the unit id, question or event description, and resolution source. Sector, salience, notional value, and outcome fields were stripped. Both coders returned all 451 units with no missing or invalid identifiers. Disagreements were adjudicated under the frozen codebook, and genuinely underdetermined units were harmonized by reviewer-final written rulings.

Cross-family Krippendorff's alpha for ordinal data was 0.921 for D1, 0.958 for D2, and 0.760 for D3 across all 451 units. In the market-only subset the corresponding values were 0.895, 0.947, and 0.748. In the event-frame subset they were 1.000, 0.997, and 0.761. The pre-specified floor was alpha >= 0.67. The final code source distribution was 349 units coded identically by both coders, 62 settled in adjudication, and 40 settled by written reviewer rulings.

The computation behind equation (2) is the standard one for ordinal data with two coders (Krippendorff 2004). The two coders' codes on the 451 units yield 902 pairable values. Let $n_c$ denote the number of times rank c appears among the pairable values, and let $o_{ck}$ denote the coincidence count, the number of times ranks c and k are paired within the same unit, summed over both orderings. The ordinal difference function weights a disagreement by the marginal mass it skips over,

$$\delta^2(c, k) = \left( \sum_{g=c}^{k} n_g - (n_c + n_k)/2 \right)^2,$$

so that a 0-versus-2 disagreement counts more heavily than a 0-versus-1 disagreement, in proportion to how the scale is actually used. Observed disagreement is $D_o = (1/n) \sum_{c,k} o_{ck} \delta^2(c, k)$ with $n = 902$, and expected disagreement $D_e$ replaces $o_{ck}$ with the chance-pairing term $n_c n_k / (n - 1)$. Alpha is then $1 - D_o / D_e$. Because $D_e$ is computed from the empirical marginals, alpha is penalized when coders agree merely by both using a dominant category, which is the property that distinguishes it from raw percent agreement.

The human benchmark protocol mattered because it was allowed to fail. In the first 48-unit blind human gold-standard check, D1 passed at alpha = 0.968 and D2 passed at alpha = 0.918, but D3 failed at alpha = 0.417. That failure forced the rewritten D3 decision tree. A fresh 24-unit D3 revalidation then passed at alpha = 0.772, with exact agreement of 83.3 percent and one 0-versus-2 disagreement. We therefore use D1 + D2 as the primary score and D1 + D2 + D3 as the secondary score.

The final D1 + D2 market distribution was: sports 3.99 (n = 80), elections 3.98 (n = 105), politics/governance 2.52 (n = 48), media mentions 2.43 (n = 7), other 2.00 (n = 13), macroeconomy 2.00 (n = 4), foreign policy 1.36 (n = 14), and security/conflict 0.67 (n = 48). The event-frame distribution was: election classes 4.00 (n = 97), national referenda 3.00 (n = 17), coup attempts 1.00 (n = 12), conflict onsets 1.00 (n = 3), and irregular leadership exits 1.00 (n = 2).

## D.3 External formation frame

The formation frame contains 131 events from January 27, 2022 through November 25, 2025. It combines 114 IFES ElectionGuide national election or referendum occasions with 17 stratum-B events: 12 coup attempts, 3 armed-conflict onsets under the UCDP episode-start rule, and 2 irregular leadership exits of chief executives. Salience is complete for all 131 events and is measured from English Wikipedia country-page views.

The boundary rules were applied symmetrically. Three excluded near misses document the rule: the Burkina Faso coup of January 24, 2022, which falls three days before the window; the DRC/M23 episode whose UCDP episode start predates the window; and the Guinea-Bissau coup of November 26, 2025, which falls one day after the window closes. The single borderline retained event is the January 8, 2023 Brasilia attack.

## D.4 Matching rules and audit trail

The matching outcome is formed = 1 if an event matched at least one paper-scope market. Elections and referenda use country, occasion, body, and timing compatibility, with first-round and runoff occasions retained separately. Point events such as coups and irregular leadership exits require country, incident or protagonist terms, and time-vicinity evidence. Conflict onsets use the v1.1 episodic correction: an onset matches any paper-scope market about the conflict episode, regardless of listing-date distance from the onset.

The v1.1 correction changed one event: the Sudan April 15, 2023 conflict onset moved from unformed to formed, adding 11 matched markets and $134,344.72 in observed value. Because Sudan is in the lowest-legibility class, this correction is conservative against the legibility hypothesis. The final matching output contains 52 formed events out of 131, 493 matched event-contract pairs, and 2,312 rejected candidate pairs retained in the near-miss audit.

### D.5 Pre-specified test rules

The primary formation test used D1 + D2 and treated the hypothesis as supported only if both criteria passed: the top-tercile formation rate had to exceed the bottom-tercile rate by at least 20 percentage points, and the legibility coefficient in the salience-controlled logit had to be positive with $p < 0.05$. A Fisher exact high-low split was pre-specified as the fallback under separation or small-cell problems. We also pre-committed to reporting the secondary D1 + D2 + D3 score, the exclusion of the single borderline event, and the elections-and-referenda-only run as sensitivity analyses.

The primary run did not meet the criteria: top-minus-bottom was 10.1 percentage points, the logit coefficient was 0.433 with $p = 0.056$, and Fisher's exact $p = 0.598$. The IFES-only run passed, but it is reported as a sensitivity rather than promoted to the headline result.

### D.6 Equations and estimation details

Equation (1) defines the primary score: $L_i = D1_i + D2_i$, with $L_i$ on a 0-to-4 scale. Equation (2) defines Krippendorff's alpha for ordinal reliability. Equation (3) defines salience as the maximum English-Wikipedia country-page view count inside a 14-day symmetric event window. Equation (4) is the listing model in which attention raises expected fee revenue and legibility lowers settlement cost. Equation (5) is the event-level logit estimated on the 131-event frame with log salience and region controls. Equation (6) is the formed-contract value regression estimated on the 269 positive-notional coded contracts, with region and sector controls and standard errors clustered on 56 topic countries; the reported robustness variant weights by observed value.

## Appendix E. Mathematical appendix

This appendix develops the estimation theory behind the formation logit (5), proves Corollary 1, reports the complete estimates for the logit and the value regressions (6) with confidence intervals, derives the marginal effects, tabulates the sensitivity runs, quantifies the power of the 131-event frame, derives the Chapman estimator interval behind Section 3.3, and develops the reliability coefficient and the salience proxy from first principles. Every number is computed from the frozen analysis outputs and the frozen event frame; no model is respecified.

### E.1 The threshold model and the estimating equations

This section derives the empirical specification (5) from the listing rule (4) and develops the estimation theory behind the reported standard errors and marginal effects. The listing rule is

$$F_e = 1\{ R(S_e) - c(L_e) + u_e \geq \kappa \},$$

with $R$ strictly increasing, $c$ strictly decreasing, and $u_e$ an idiosyncratic shock with cumulative distribution function $G$, independent of $(L_e, S_e)$. Proposition 1, proved in Section 6.1, states that the implied listing probability $\Pr(F_e = 1 \mid l, s) = 1 - G(\kappa - R(s) + c(l))$ is increasing in both arguments.

**From the threshold model to the logit.** Assume the shock follows the standard logistic distribution, $G(t) = 1/(1 + e^{-t})$, whose density $g(t) = G(t)(1 - G(t))$ is symmetric and log-concave. By symmetry, $1 - G(t) = \Lambda(-t)$ where $\Lambda$ is the logistic function, so

$$\Pr(F_e = 1 \mid L_e, S_e) = \Lambda( R(S_e) - c(L_e) - \kappa ).$$

Specifying revenue as log-linear in attention, $R(s) = \beta_2 \ln s$, settlement cost as linear in legibility, $-c(l) = \beta_1 l$ with $\beta_1 > 0$ capturing $c' < 0$, absorbing $-\kappa$ into the intercept $\beta_0$, and adding the region shifter $\beta_3$ $LatAm_e$ to allow the latent listing index to differ by region, gives exactly the estimating equation (5),

$$\Pr(F_e = 1) = \Lambda( \beta_0 + \beta_1 L_e + \beta_2 \ln S_e + \beta_3 LatAm_e ) = \Lambda(x_e'\beta),$$

where $x_e$ collects the regressors. The logit is therefore not a generic link choice; it is the threshold model under a logistic shock, and the comparative statics of Proposition 1 are the sign predictions $\beta_1 > 0$ and $\beta_2 > 0$.

**Likelihood, score, and information.** With independent events, the log-likelihood is

$$\ell(\beta) = \Sigma_e [ F_e \ln \Lambda(x_e'\beta) + (1 - F_e) \ln(1 - \Lambda(x_e'\beta)) ].$$

Using $\Lambda' = \Lambda(1 - \Lambda)$, the score vector is

$$\partial\ell/\partial\beta = \Sigma_e ( F_e - \Lambda(x_e'\beta) ) x_e,$$

so the maximum-likelihood estimate $\hat{\beta}$ solves the orthogonality condition $\Sigma_e (F_e - \hat{\Lambda}_e) x_e = 0$: residuals are uncorrelated with every regressor, including the intercept, which forces the average fitted probability to equal the observed formation rate of 52/131. The observed information matrix is

$$I(\beta) = \Sigma_e \Lambda_e (1 - \Lambda_e) x_e x_e',$$

which is also the expected information because the logit score has conditional variance $\Lambda(1 - \Lambda)$. Estimation proceeds by Newton iteration on the score, and the standard errors reported in Table E.1 are the square roots of the diagonal of $I(\hat{\beta})^{-1}$, the inverse-Hessian estimator. With $n = 131$ and 52 successes, the effective sample is small; the p-values therefore rely on the asymptotic normality of $\hat{\beta}$, which is why a Fisher exact fallback was pre-specified.

**Marginal effects.** The marginal effect of legibility at covariate vector x is

$$\partial\Pr(F = 1 \mid x)/\partial L = \beta_1 \lambda(x'\beta),\ \lambda(t) = \Lambda(t)(1 - \Lambda(t)),$$

evaluated at the sample mean $\bar{x}$ for the effect at means and averaged over the empirical distribution of $x_e$ for the average marginal effect; Appendix E.4 reports both. For inference on the effect at means, the delta method gives

$$\mathrm{Var}( \hat{\beta}_1 \lambda(\bar{x}'\hat{\beta}) ) = \nabla h(\hat{\beta})' I(\hat{\beta})^{-1} \nabla h(\hat{\beta}),\ \nabla h(\beta) = \lambda(\bar{x}'\beta) [ e_1 + \beta_1 (1 - 2\Lambda(\bar{x}'\beta)) \bar{x} ],$$

where $e_1$ selects the legibility coordinate. Because $\lambda(t) \le 1/4$ everywhere, the marginal effect of any coefficient is bounded by a quarter of the coefficient, which is the familiar logit quarter rule; at the observed means the fitted probability is close enough to one half that the rule is nearly exact, giving $0.433/4 \approx 0.108$ against the exact 0.103.

## E.2 Corollary 1: statement and proof

The corollary requires one regularity condition beyond (4): the shock density must be log-concave. The logistic density assumed in (5) satisfies it, as do the normal, extreme-value, and most textbook shock distributions.

**Corollary 1.** *Suppose $L_e$ and $S_e$ are independent in the population, $u_e$ is independent of $(L_e, S_e)$, and $u_e$ has a log-concave density g. Then, conditional on $F_e = 1$, $L_e$ and $S_e$ are negatively dependent: for every pair of increasing functions $\varphi$ and $\psi$,*

$$\mathrm{Cov}(\ \varphi(L_e), \psi(S_e) \mid F_e = 1\ ) \leq 0,$$

*with strict inequality when g is strictly log-concave and $L_e$, $S_e$ are non-degenerate.*

*Proof.* Write $a = R(S_e)$ and $b = -c(L_e)$. Both are increasing transformations, so a is increasing in $S_e$ and b is increasing in $L_e$, and it suffices to prove the covariance inequality for (a, b). Let $\bar{G}(t) = 1 - G(t)$ denote the survival function of $u_e$. Because g is log-concave, $\bar{G}$ is log-concave as well, a standard property of log-concave densities; hence $(\ln \bar{G})'$ is non-increasing.

By Proposition 1, $\Pr(F_e = 1 \mid a, b) = \bar{G}(\kappa - a - b)$. With a and b independent in the population with densities $f_a$ and $f_b$, Bayes' rule gives the joint density conditional on listing,

$$f(a, b \mid F = 1) \propto f_a(a)\ f_b(b)\ \bar{G}(\kappa - a - b).$$

Consider the conditional density of b given a and F = 1, and take any $a' > a$. The likelihood ratio

$$f(b \mid a', F = 1)\ /\ f(b \mid a, F = 1) \propto \bar{G}(\kappa - a' - b)\ /\ \bar{G}(\kappa - a - b)$$

has logarithmic derivative in b equal to $(\ln \bar{G})'(\kappa - a - b) - (\ln \bar{G})'(\kappa - a' - b)$. Since $\kappa - a' - b < \kappa - a - b$ and $(\ln \bar{G})'$ is non-increasing, this derivative is non-positive: the ratio is non-increasing in b. The conditional law of b given a, F = 1 is therefore decreasing in a in the likelihood-ratio order, which implies first-order stochastic dominance, so $E[\tilde{\psi}(b) \mid a, F = 1]$ is non-increasing in a for every increasing $\tilde{\psi}$. By the law of total covariance applied conditional on F = 1,

$$\mathrm{Cov}(\ \tilde{\varphi}(a), \tilde{\psi}(b) \mid F = 1\ ) = \mathrm{Cov}(\ \tilde{\varphi}(a), E[\tilde{\psi}(b) \mid a, F = 1] \mid F = 1\ ) \leq 0,$$

because the covariance of an increasing and a non-increasing function of the same random variable is non-positive. Strict log-concavity makes the likelihood-ratio monotonicity strict and the inequality strict. ■

The economic content is the sentence in the main text. Selection truncates the set $\{a + b + u \geq \kappa\}$: among survivors, a low value of one argument must be compensated by a high value of another, so an illegible event appears in the inventory only when its salience is extreme. This is the standard endogenous-selection logic of conditioning on a collider (Heckman 1979; Elwert and Winship 2014), specialized to the threshold structure of (4), and it is why Section 6.4 treats the negative value coefficient as confirmation of the selection model rather than as evidence against the formation mechanism.

### E.3 Formation logit: complete estimates

Table E.1 reports every coefficient of the logit (5) for the pre-registered primary run and the three pre-committed sensitivity runs. Standard errors are inverse-Hessian; confidence intervals are coefficient ± 1.96 standard errors.

Table E.1. Logit estimates, Pr(F_e = 1), n as stated per run

| Run | Term | Coefficient | SE | z | p | 95% CI |
|---|---|---|---|---|---|---|
| Primary, full frame (n = 131, 52 formed) | Intercept | −6.989 | 2.808 | −2.49 | 0.013 | [−12.49, −1.49] |
| | Legibility L (D1+D2) | 0.433 | 0.226 | 1.91 | 0.056 | [−0.01, 0.88] |
| | ln salience | 0.485 | 0.248 | 1.96 | 0.050 | [−0.00, 0.97] |
| | Latin America | 1.030 | 0.380 | 2.71 | 0.007 | [0.29, 1.77] |
| Excluding borderline event (n = 130) | Intercept | −6.866 | 2.793 | −2.46 | 0.014 | [−12.34, −1.39] |
| | Legibility L (D1+D2) | 0.390 | 0.228 | 1.71 | 0.087 | [−0.06, 0.84] |
| | ln salience | 0.488 | 0.247 | 1.98 | 0.048 | [0.01, 0.97] |
| | Latin America | 1.059 | 0.381 | 2.78 | 0.005 | [0.31, 1.81] |
| Secondary 0–6 score (n = 131) | Intercept | −5.697 | 2.603 | −2.19 | 0.029 | [−10.80, −0.60] |
| | Legibility D1+D2+D3 | 0.184 | 0.128 | 1.43 | 0.153 | [−0.07, 0.44] |
| | ln salience | 0.411 | 0.238 | 1.72 | 0.085 | [−0.06, 0.88] |
| | Latin America | 0.985 | 0.378 | 2.60 | 0.009 | [0.24, 1.73] |
| Elections and referenda only (n = 114, sensitivity) | Intercept | −13.151 | 4.358 | −3.02 | 0.003 | [−21.69, −4.61] |
| | Legibility L (D1+D2) | 2.290 | 0.819 | 2.80 | 0.005 | [0.68, 3.90] |
| | ln salience | 0.357 | 0.287 | 1.24 | 0.213 | [−0.21, 0.92] |
| | Latin America | 1.173 | 0.431 | 2.72 | 0.006 | [0.33, 2.02] |

### E.4 Marginal effects

Marginal effects for the primary run are evaluated from the Table E.1 coefficients at the frame covariate means (mean L = 3.48, mean ln salience = 9.41, Latin America share 0.40). The marginal effect of legibility at the means is 0.103, and the discrete effect of a one-point increase in L from the mean is 0.106; this is the basis for the statement in Section 6.3 that one point of legibility corresponds to roughly 10 percentage points of formation probability. The average marginal effect over the 131 frame events is 0.092. The corresponding values for ln salience are 0.115 at the means and 0.103 averaged over events.

## E.5 Value regressions: complete estimates

Table E.2 reports every coefficient of the value regression (6) on the 269 positive-notional coded contracts, with standard errors clustered on 56 topic countries. The omitted sector category is elections; the omitted region is Africa.

Table E.2. ln observed value among formed contracts (n = 269, 56 clusters)

| Term | OLS | Value-weighted OLS |
|---|---:|---:|
| Legibility L (D1+D2) | −0.782 (0.354), p = 0.027, [−1.47, −0.09] | −0.489 (0.165), p = 0.003, [−0.81, −0.17] |
| Latin America | 0.916 (0.498), p = 0.066, [−0.06, 1.89] | 1.567 (0.401), p < 0.001, [0.78, 2.35] |
| Entertainment/culture | −1.615 (0.599), p = 0.007, [−2.79, −0.44] | −4.396 (0.528), p < 0.001, [−5.43, −3.36] |
| Foreign policy/diplomacy | −3.229 (1.280), p = 0.012, [−5.74, −0.72] | −3.456 (0.703), p < 0.001, [−4.83, −2.08] |
| Macroeconomy | −0.106 (0.820), p = 0.897, [−1.71, 1.50] | −1.774 (0.392), p < 0.001, [−2.54, −1.01] |
| Other | 0.297 (1.473), p = 0.840, [−2.59, 3.18] | −0.061 (0.520), p = 0.906, [−1.08, 0.96] |
| Political media mentions | −2.267 (1.122), p = 0.043, [−4.47, −0.07] | −1.584 (0.386), p < 0.001, [−2.34, −0.83] |
| Politics/governance | −2.159 (0.857), p = 0.012, [−3.84, −0.48] | −2.160 (0.567), p < 0.001, [−3.27, −1.05] |
| Security/conflict | −0.929 (1.277), p = 0.467, [−3.43, 1.57] | −0.555 (0.705), p = 0.431, [−1.94, 0.83] |
| Sports | −2.571 (0.824), p = 0.002, [−4.19, −0.96] | −1.121 (0.762), p = 0.141, [−2.62, 0.37] |
| Intercept | 13.556 (1.532), p < 0.001, [10.55, 16.56] | 15.460 (0.939), p < 0.001, [13.62, 17.30] |

Cells report coefficient (clustered SE), p-value, and 95 percent confidence interval. The rank-based companion statistics are a Spearman correlation of −0.247 ($p < 0.001$) over all 320 coded contracts and a value-weighted Spearman correlation of −0.318 ($p < 0.001$) over the 269 positive-notional contracts.

## E.6 Sensitivity runs tabulated

Table E.3 collects the pre-committed sensitivity runs of the formation test against both pre-specified criteria. C1 requires a top-minus-bottom tercile gap of at least 20 percentage points; C2 requires a positive legibility coefficient at $p < 0.05$ in the salience-controlled logit.

Table E.3. Formation-test runs against the pre-specified criteria

| Run | n | Formed | Tercile gap (pp) | $\beta_1$ | p | Fisher p | Verdict |
|---|---|---|---|---|---|---|---|
| Primary, full frame | 131 | 52 | +10.1 | 0.433 | 0.056 | 0.598 | Not accepted |
| Excluding borderline event | 130 | 52 | +7.9 | 0.390 | 0.087 | 0.599 | Not accepted |
| Secondary 0–6 score | 131 | 52 | +5.8 | 0.184 | 0.153 | 0.792 | Not accepted |
| Elections and referenda only | 114 | 46 | +33.6 | 2.290 | 0.005 | 0.014 | Passes; reported as sensitivity only |

### E.7 Power and minimum detectable effects

The phrase "powered to detect large effects only" in Section 8 has the following quantitative content.

For criterion C1, the tercile comparison sets 17 bottom-tercile events against 97 top-tercile events. At the observed bottom-tercile rate of 35.3 percent, a two-sided two-proportion test at the 5 percent level has 32 percent power against a true 20-point gap, the gap the criterion requires, and reaches 80 percent power only at a true gap of about 35 points. The pre-registered bar therefore sits well inside the region where a true effect of exactly criterion size would usually fail to reject.

For criterion C2, power is computed by simulation over the frozen frame covariates: outcomes are redrawn from the model (5) at hypothetical values of $\beta_1$, with the other coefficients held at their estimates and the intercept recalibrated so the mean formation rate stays at the observed 52/131, and the estimation logit is refit to each draw (2,000 draws per value). The power to detect $\beta_1$ at $p < 0.05$ is 14 percent at $\beta_1 = 0.2$, 52 percent at the observed 0.433, 75 percent at 0.6, 94 percent at 0.8, and 99 percent at 1.0. The design reaches conventional 80 percent power at roughly $\beta_1 \approx 0.65$, one and a half times the observed coefficient.

Two implications follow. A frame of this size could only have passed C2 reliably if the formation effect were substantially larger than the one estimated, and the observed near-miss at $p = 0.056$ is the expected outcome when the true effect equals the estimate, since power at that value is close to a coin flip. Neither implication rescues the hypothesis; both calibrate what the non-rejection does and does not say.

### E.8 Chapman estimator detail

Section 3.3 uses the Chapman bias-corrected form of the Lincoln-Petersen capture-recapture estimator. With $n_A = 49$ markets flagged by the keyword pass, $n_B = 56$ flagged by the semantic pass, and $m = 30$ flagged by both, the point estimate is

$$\hat{N} = (n_A + 1)(n_B + 1)/(m + 1) - 1 = (50 \times 57)/31 - 1 = 90.94.$$

Its estimated variance is the standard Chapman form,

$$\mathrm{Var}(\hat{N}) = (n_A + 1)(n_B + 1)(n_A - m)(n_B - m) / [ (m + 1)^2 (m + 2) ]$$
$= (50 \times 57 \times 19 \times 26) / (31^2 \times 32) = 45.78$,

giving a standard error of 6.77 and a normal-approximation 95 percent interval of $90.94 \pm 1.96 \times 6.77 = [77.7, 104.2]$. Dividing by the sample size of 2,000 converts these to the prevalence figures in the text: a point estimate of 4.5 percent with interval 3.9 to 5.2 percent. As Section 3.3 notes, positive dependence between the two classifiers biases $\hat{N}$ downward, which is why the exercise is read as a diagnostic on the failure mode rather than as a population total.

### E.9 The reliability coefficient: derivation and properties

This section develops the reliability coefficient of equation (2) from first principles and establishes the properties that justify its use, so that the reported alphas can be read without reference to outside sources.

**Setup.** Two coders each assign one of the ordinal ranks {0, 1, 2} to each of the $N = 451$ units on a given dimension, yielding $n = 2N = 902$ pairable values. Let $n_c$ be the number of times rank c appears among the pairable values, and define the coincidence count $o_{ck}$ as the number of ordered within-unit pairs in which one coder assigned c and the other assigned k, so that $\Sigma_c \Sigma_k o_{ck} = n$ and the matrix o is symmetric. The ordinal difference function is

$$\delta^2(c, k) = \left( \Sigma_{g=c}^{k} n_g - (n_c + n_k)/2 \right)^2,$$

the squared marginal mass lying between the two ranks. Observed and expected disagreement are

$$D_o = (1/n) \Sigma_c \Sigma_k o_{ck} \delta^2(c, k), \quad D_e = (1/n) \Sigma_c \Sigma_{k \neq c} ( n_c n_k / (n - 1) ) \delta^2(c, k),$$

and $\alpha = 1 - D_o / D_e$.

**Where the expected-disagreement term comes from.** $D_e$ is not an arbitrary normalizer; it is $D_o$ computed under random pairing. Pool all n values and draw an ordered pair without replacement. The probability of drawing ranks (c, k) with $c \neq k$ is $n_c n_k / (n(n - 1))$, so the expected value of $\delta^2$ under random pairing is exactly the $D_e$ expression. Alpha therefore compares the disagreement coders actually produced with the disagreement that the same marginal distribution of codes would produce if codes were shuffled across units.

**Property 1 (perfect agreement).** $\alpha = 1$ if and only if the coders agree on every unit. If they agree everywhere, every within-unit pair has $c = k$, $\delta^2(c, c) = 0$, so $D_o = 0$ and $\alpha = 1$. Conversely, if any unit is coded (c, k) with $c \neq k$, then since the marginal masses $n_c$ and $n_k$ are positive for those ranks, $\delta^2(c, k) > 0$ and $D_o > 0$, so $\alpha < 1$. ■

**Property 2 (chance agreement).** If codes are paired at random, in the sense that the within-unit pairing is distributed as a random draw from the pooled marginals, then $E[D_o] = D_e$ up to the without-replacement correction embedded in both terms, so $E[\alpha] = 0$. Agreement that merely reflects both coders using a dominant category is thus credited as chance, not as reliability. This is the property that separates alpha from raw percent agreement: two coders who both assign the modal rank to every unit can reach high percent agreement while alpha is near zero, because $D_e$ shrinks with the concentration of the marginals exactly as $D_o$ does. ■

**Property 3 (ordinal weighting).** The metric $\delta^2$ weights a disagreement by the squared empirical mass between the two ranks, so a 0-versus-2 disagreement always counts at least as much as a 0-versus-1 disagreement, with the exact ratio determined by how the scale is used in the data rather than by an assumption of equal spacing. The metric is invariant to order-preserving relabelings of the ranks, which is the correct invariance class for an ordinal instrument: nothing in the analysis depends on the codes being the numerals 0, 1, 2 rather than any other increasing labels.

**Range and the pre-registered floor.** Alpha is bounded above by 1, equals 0 at chance, and is negative when coders disagree more systematically than chance would produce. The pre-registered floor $\alpha \geq 0.67$ is Krippendorff's (2004) conventional threshold for drawing tentative conclusions; all production values in Appendix D.2 clear it, and the one benchmark value that did not, the first-stage D3 check at 0.417, was treated as the failure it was.

### E.10 The salience proxy

The salience control in equation (3) is

$$S_e = \max \text{ over } t \in [t_e - 14, t_e + 14] \text{ of } v_c(t),$$

where $v_c(t)$ is the daily English-Wikipedia pageview count of the country page for the country of event e and $t_e$ is the event day. Three design properties motivate the form, all fixed before estimation. First, the window is symmetric because the frame mixes scheduled events, whose attention is anticipatory and peaks before $t_e$, with unscheduled events such as coups, whose attention is reactive and peaks after; a symmetric window treats both classes identically rather than building the event class into the control. Second, the maximum rather than the mean is used because the listing decision responds to whether an uncertainty commands attention at all, and the maximum captures the attention peak invariantly to where in the window it falls: for any two events whose attention spikes lie inside the window, the proxy compares the spikes themselves, not their timing. Third, the regression uses $\ln S_e$ because daily pageview maxima are heavy-tailed across countries with very different baseline traffic, and the logarithm makes the salience coefficient an elasticity-type effect rather than one dominated by the largest countries. The window half-width of 14 days is a design constant, not a tuned parameter; it was fixed when the frame was frozen, before any market data were matched.

## References

Arrow, Kenneth J., Robert Forsythe, Michael Gorham, Robert Hahn, Robin Hanson, John O. Ledyard, Saul Levmore, Robert Litan, Paul Milgrom, Forrest D. Nelson, George R. Neumann, Marco Ottaviani, Thomas C. Schelling, Robert J. Shiller, Vernon L. Smith, Erik Snowberg, Cass R. Sunstein, Paul C. Tetlock, Philip E. Tetlock, Hal R. Varian, Justin Wolfers, and Eric Zitzewitz. 2008. "The Promise of Prediction Markets." *Science* 320(5878): 877-878. https://doi.org/10.1126/science.1157679.

Berg, Joyce E., Forrest D. Nelson, and Thomas A. Rietz. 2008. "Prediction Market Accuracy in the Long Run." *International Journal of Forecasting* 24(2): 285-300. https://doi.org/10.1016/j.ijforecast.2008.03.007.

Bürgi, Constantin, Wanying Deng, and Karl Whelan. 2025. "Makers and Takers: The Economics of the Kalshi Prediction Market." CESifo Working Paper No. 12122. https://ssrn.com/abstract=5502658.

Çalışkan, Koray, and Michel Callon. 2009. "Economization, Part 1: Shifting Attention from the Economy Towards Processes of Economization." *Economy and Society* 38(3): 369-398. https://doi.org/10.1080/03085140903020580.

Callon, Michel, ed. 1998. *The Laws of the Markets*. Oxford: Blackwell.

Eisensee, Thomas, and David Strömberg. 2007. "News Droughts, News Floods, and U.S. Disaster Relief." *Quarterly Journal of Economics* 122(2): 693-728. https://doi.org/10.1162/qjec.122.2.693.

Elwert, Felix, and Christopher Winship. 2014. "Endogenous Selection Bias: The Problem of Conditioning on a Collider Variable." *Annual Review of Sociology* 40: 31-53. https://doi.org/10.1146/annurev-soc-071913-043455.

Gebele, Jonas, and Florian Matthes. 2026. "Semantic Non-Fungibility and Violations of the Law of One Price in Prediction Markets." arXiv:2601.01706. https://arxiv.org/abs/2601.01706.

Gilardi, Fabrizio, Meysam Alizadeh, and Maël Kubli. 2023. "ChatGPT Outperforms Crowd Workers for Text-Annotation Tasks." *Proceedings of the National Academy of Sciences* 120(30): e2305016120. https://doi.org/10.1073/pnas.2305016120.

Grossman, Sanford J., and Oliver D. Hart. 1986. "The Costs and Benefits of Ownership: A Theory of Vertical and Lateral Integration." *Journal of Political Economy* 94(4): 691-719. https://doi.org/10.1086/261404.

Hanson, Robin. 2007. "Logarithmic Market Scoring Rules for Modular Combinatorial Information Aggregation." *Journal of Prediction Markets* 1(1): 3-15.

Hart, Oliver. 2017. "Incomplete Contracts and Control." *American Economic Review* 107(7): 1731-1752. https://doi.org/10.1257/aer.107.7.1731.

Hayek, F. A. 1945. "The Use of Knowledge in Society." *American Economic Review* 35(4): 519-530.

Heckman, James J. 1979. "Sample Selection Bias as a Specification Error." *Econometrica* 47(1): 153-161. https://doi.org/10.2307/1912352.

Jia, Huaiyu, Luofeng Zhou, Wentao Zhang, Lin William Cong, Siguang Li, and Shuo Sun. 2026. "Unlocking the Forecasting Economy: A Suite of Datasets for the Full Lifecycle of Prediction Market: [Experiments & Analysis]." arXiv:2604.20421. https://arxiv.org/abs/2604.20421.

Krippendorff, Klaus. 2004. *Content Analysis: An Introduction to Its Methodology*. 2nd ed. Thousand Oaks, CA: Sage.

Le, Nam Anh. 2026. "Decomposing Crowd Wisdom: Domain-Specific Calibration Dynamics in Prediction Markets." arXiv:2602.19520. https://arxiv.org/abs/2602.19520.

MacKenzie, Donald, and Yuval Millo. 2003. "Constructing a Market, Performing Theory: The Historical Sociology of a Financial Derivatives Exchange." *American Journal of Sociology* 109(1): 107-145. https://doi.org/10.1086/374404.

Madhavan, Ananth. 2000. "Market Microstructure: A Survey." *Journal of Financial Markets* 3(3): 205-258. https://doi.org/10.1016/S1386-4181(00)00007-0.

McCarthy, John D., Clark McPhail, and Jackie Smith. 1996. "Images of Protest: Dimensions of Selection Bias in Media Coverage of Washington Demonstrations, 1982 and 1991." *American Sociological Review* 61(3): 478-499. https://doi.org/10.2307/2096360.

Reichenbach, Felix, and Martin Walther. 2025. "Exploring Decentralized Prediction Markets: Accuracy, Skill, and Bias on Polymarket." SSRN. https://ssrn.com/abstract=5910522.

Rhode, Paul W., and Koleman S. Strumpf. 2013. "The Long History of Political Betting Markets: An International Perspective." In *The Oxford Handbook of the Economics of Gambling*, edited by Leighton Vaughan Williams and Donald S. Siegel, 559-586. Oxford: Oxford University Press. https://doi.org/10.1093/oxfordhb/9780199797912.013.0029.

Rohanifar, Yasaman, Syed Ishtiaque Ahmed, and Sharifa Sultana. 2026. "Prediction Laundering: The Illusion of Neutrality, Transparency, and Governance in Polymarket." arXiv:2602.05181. https://arxiv.org/abs/2602.05181.

Roth, Alvin E. 2007. "Repugnance as a Constraint on Markets." *Journal of Economic Perspectives* 21(3): 37-58. https://doi.org/10.1257/jep.21.3.37.

Scott, James C. 1998. *Seeing Like a State: How Certain Schemes to Improve the Human Condition Have Failed*. New Haven, CT: Yale University Press.

Silber, William L. 1981. "Innovation, Competition, and New Contract Design in Futures Markets." *Journal of Futures Markets* 1(2): 123-155. https://doi.org/10.1002/fut.3990010205.

Snowberg, Erik, Justin Wolfers, and Eric Zitzewitz. 2013. "Prediction Markets for Economic Forecasting." In *Handbook of Economic Forecasting*, vol. 2A, edited by Graham Elliott and Allan Timmermann, 657-687. Amsterdam: Elsevier. https://doi.org/10.1016/B978-0-444-53683-9.00011-6.

Törnberg, Petter. 2023. "ChatGPT-4 Outperforms Experts and Crowd Workers in Annotating Political Twitter Messages with Zero-Shot Learning." arXiv:2304.06588. https://arxiv.org/abs/2304.06588.

Wolfers, Justin, and Eric Zitzewitz. 2004. "Prediction Markets." *Journal of Economic Perspectives* 18(2): 107-126. https://doi.org/10.1257/0895330041371321.

Wolfers, Justin, and Eric Zitzewitz. 2006. "Prediction Markets in Theory and Practice." NBER Working Paper No. 12083. https://doi.org/10.3386/w12083.

Ziems, Caleb, William Held, Omar Shaikh, Jiaao Chen, Zhehao Zhang, and Diyi Yang. 2024. "Can Large Language Models Transform Computational Social Science?" *Computational Linguistics* 50(1): 237-291. https://doi.org/10.1162/coli_a_00502.